\documentclass{amsart}
\usepackage{latexsym}
\usepackage{amsfonts}

\newtheorem{theorem}{Theorem}[section]
\newtheorem{lemma}[theorem]{Lemma}
\newtheorem{corollary}[theorem]{Corollary}
\newtheorem{proposition}[theorem]{Proposition}
\newtheorem{remark}[theorem]{Remark}

\theoremstyle{definition}

\newtheorem{example}[theorem]{Example}

\def\nn{\nonumber}
\def\d{\hbox{\rm d}}

\def\ifrac#1#2{{#1}/{(#2)}}
\def\od#1{\frac{\d}{\d#1}}
\def\pd#1{\frac{\partial}{\partial#1}}
\def\pderiv#1#2{\frac{\partial#1}{\partial#2}}
\def\hide#1{}
\theoremstyle{remark}

\newcommand{\res}{{\rm Res}}

\newcommand{\tr}{{\rm tr}}

\newcommand{\p}{\partial}

\theoremstyle{remark}

\begin{document}
\def\tr{{\rm Tr}}
\def\bk{B\"acklund}
\def\p{\partial}
\def\Pa{Painlev\'e}
\def\Pt{{\rm P}_{\rm{\scriptstyle II}}}
\def\Ptn{{\rm P}^{(n)}_{\rm{\scriptstyle II}}}
\def\Lr{\mathcal{L}}
\def\Pol{{\mathcal P}}
\def\Se{{\mathcal S}}
\def\A{{\mathcal A}}
\def\M{{\mathsf M}}
\def\Complex{{\mathbb C}}
\def\Real{{\mathbb R}}
\def\ID{\mathbb{I}}
\def\Integer{{\mathbb Z}}
\def\ddt{{{\rm d}\over{\rm d} t}}
\def\ddx{{{\rm d}\over{\rm d} x}}
\def\ddtt{{{\rm d}\over{\rm d}\tau}}
\def\ddz{{{\rm d}\over{\rm d} z}}
\def\thi{{\vartheta_\infty}}
\def\eps{\varepsilon}
\def\arg{{\rm arg}}
\def\iff{\Longleftrightarrow}
\def\U{{\mathcal U}}

\title[Hamiltonian structure of the PII Hierarchy]
{The Hamiltonian Structure of the Second Painlev\'e Hierarchy.}

\author{Marta Mazzocco}
\noindent\address{\noindent School of Mathematics, The University
of Manchester, Manchester M60 1QD, United Kingdom}
\email{Marta.Mazzocco@manchester.ac.uk}
\author{Man Yue Mo}
\noindent\address{\noindent Department of Mathematics,  University of Bristol,
Bristol BS8 1TW, United Kingdom}
\email{M.Mo@bristol.ac.uk}


\maketitle

\begin{abstract}
In this paper we study the Hamiltonian structure of the second Painlev\'e hierarchy,
an infinite sequence of nonlinear ordinary differential equations containing PII as its simplest equation.
The $n$-th element of the hierarchy is a non linear ODE of order $2n$ in the independent variable $z$ depending on $n$ parameters denoted by ${t}_1,\dots,{t}_{n-1}$ and $\alpha_n$. We introduce new canonical coordinates and obtain Hamiltonians for the $z$ and $t_1,\dots,t_{n-1}$ evolutions. We give explicit formulae for these Hamiltonians showing that they are polynomials in our canonical coordinates.
\end{abstract}

\tableofcontents

\section{Introduction}

In this paper we study the Hamiltonian structure of the second Painlev\'e hierarchy,
an infinite sequence of nonlinear ordinary differential equations containing
$$
\Pt :\quad\quad w_{zz} = 2 w^3+z\,w+\alpha_1 ,
$$
as its simplest equation.
The $n$-th element of the hierarchy is a non linear ODE of order $2n$, depending on $n$ parameters denoted by ${t}_1,\dots,{t}_{n-1}$ and $\alpha_n$:
$$
 \Ptn :\quad\quad
\left(\od{z}+2 w\right) \Lr_{n}\left[w_{z}-w^{2}\right]
+\sum_{l=1}^{n-1} {t}_{l}\left(\od{z}+2 w\right)
 \Lr_{l}\left[w_{z}-w^{2}\right]
=z w+\alpha_{n},\qquad n\ge 1,
$$
where $\Lr_{n}$ is the
operator defined by the recursion relation
\begin{equation}\label{i:lenard}
\frac{{\rm d}}{{\rm d}z}
{\mathcal L}_{n+1}= \left(\frac{{\rm d^3}}{{\rm d}z^3}+4(w_z-w^2)
\frac{{\rm d}}{{\rm d}z}+2(w_z-w^2)_z\right){{\mathcal L}}_n,
\qquad {\mathcal L}_0[w_z-w^2] = {\textstyle\frac12}.
\end{equation}

The second Painlev\'e equation and its hierarchy appear in several applications
including Hele-Shaw geometry
\cite{FT}, nonlinear optics \cite{GJ} and random matrix theory \cite{TW,CKV} to name only a few.

The Hamiltonian structure of the classical six Painlev\'e equations was discovered long ago by
Okamoto \cite{okam}, Jimbo and Miwa \cite{JM}. In the case of $n=1$, i.e.  $\Pt$, the Hamiltonian is
$$
\mathcal H^{(1)}= 4 P^2 +\frac{1}{4} Q+ \frac{1}{4} P Q^2+2 P z-\frac{1}{2} Q \alpha_1.
$$
where
$$
Q=4 w,\qquad P=\frac{1}{2}\left(w_z-w^2-\frac{z}{2}\right).
$$
Using such formulation Okamoto was able to describe his initial conditions space, to characterize the
action of the B\"ack\-lund transformations found by Gambier \cite{gam} and Lukashevich \cite{luk}
in terms of affine Weyl groups and to produce immediately the so called Riccati-type classical solutions of the second Painelv\'e equation \cite{gam,gro,refAirault}.  Also several properties of the Yablonskii-Vorob'ev polynomials  \cite{Ya,Vo}  describing the rational solutions were proved using the Hamiltonian formulation \cite{CM}.
Umemura and Watanabe \cite{UW} used  the Hamiltonian structure in to prove the irreducibility of $\Pt$.

In this paper we introduce canonical coordinates $P_1,\dots,P_n,Q_1,\dots,Q_n$ and a Hamiltonian
function ${\mathcal H}^{(n)}$ such that $\Ptn$ is equivalent to
\begin{equation}\label{i:hpq}
\frac{\partial Q_i}{\partial z}=\frac{\partial {\mathcal H}^{(n)}}{\partial P_i},
\qquad
\frac{\partial P_i}{\partial z}=-\frac{\partial {\mathcal H}^{(n)}}{\partial Q_i},
\quad i=1,\dots,n.
\end{equation}
In particular we show that $ {\mathcal H}^{(n)}$ is a polynomial in
$P_1,\dots,P_n,Q_1,\dots,Q_n$ and that the Hamiltonian equations
satisfy the Painlev\'e property.

Observe that starting from the second element of the hierarchy, the
parameters ${t}_1,\dots,{t}_{n-1}$ appear in $\Ptn$. The solutions
$w$ will depend on the {\it times} ${t}_1,\dots,{t}_{n-1}$ according
to the equation
\begin{equation}\label{i:tf}
(2k+1)\frac{\p w}{\p {t}_k} + \p_z\left(\p_{z}+
2 w\right) \Lr_{k}\left[w_{z}-w^{2}\right]=0,\qquad k=1,\dots,n-1,
\end{equation}
which is actually the $k$-th element of the  mKdV hierarchy. In fact, the second Painlev\'e hierarchy was discovered as self--similarity reduction of the mKdV hierarchy \cite{refAS77,refAirault,FN} (details on this derivation are recalled in section \ref{se:pii} below).

We refer to the evolution in the times
${t}_1,\dots,{t}_{n-1}$, as {\it time-flows.}\/ We prove that the time-flows are Hamiltonian and we compute  the Hamiltonians ${\mathcal H}^{(n)}_1,\dots,{\mathcal H}^{(n)}_{n-1}$ such that the system
$$
\frac{\partial Q_i}{\partial{t}_k}=\frac{\partial  {\mathcal H}^{(n)}_k}{\partial P_i},
\qquad
\frac{\partial P_i}{\partial{t}_k}=-\frac{\partial  {\mathcal H}^{(n)}_k}{\partial Q_i},
\quad i=1,\dots,n,
$$
is equivalent to (\ref{i:tf}). These Hamiltonians are also polynomials in $P_1,\dots,P_n,Q_1,\dots,Q_n$ and again the Painlev\'e property is satisfied. Amazingly, we obtain explicit formulae for
${\mathcal H}^{(n)}$ and $\mathcal H_k^{(n)}$, $k=1,\dots,n-1$, in terms of $P_1,\dots,P_n,Q_1,\dots,Q_n$ without need of recursion.

Our construction is based on the fact that the second Painlev\'e hierarchy can be interpreted as monodromy preserving deformation equation\footnote{The isomonodromic deformation problem for the second Painlev\'e hierarchy with
${t}_1=\dots={t}_{n-1}=0$ was derived in \cite{refCJM} and in \cite{Kud} following
the approach proposed in \cite{AKNS} starting form the isomonodromy deformation problem given in \cite{FN} for the second Painlev\'e equation. Here we generalize the
construction of  \cite{refCJM}  to the case of generic values ${t}_1,\dots, {t}_{n-1}$.} of an auxiliary linear system
$$
\frac{{\rm d} \Psi}{{\rm d}\lambda} = \mathcal A^{(n)}(\lambda;z,{t}_1,\dots,{t}_{n-1}) \Psi
$$
where $ \mathcal A^{(n)}$ is a matrix function of $\lambda$ holomorphic in $\mathbb C^\ast$, having a simple pole at $0$ and a pole of order $2n+2$ at infinity. The isomonodromic condition is expressed by
the zero--curvature conditions
\begin{subequations}
\begin{eqnarray}
&&
\label{i:comp-ii}
\frac{\partial{\mathcal A}^{(n)}}{\partial z}  -\frac{\partial{\mathcal B}}{\partial \lambda}=
[{\mathcal B},{\mathcal A}^{(n)}], \qquad {\mathcal B}=
-\left({{\mathcal A}^{(n)}\lambda^{1-2n}\over 4^{n}}\right)_+,\\
&&
\label{i:laxMA}
\p_{\lambda}\hat M^{(k)}-(2k+1)\p_{{t}_k}\mathcal{A}^{(n)}=-[\hat M^{(k)},\mathcal{A}^{(n)}],
\end{eqnarray}
\end{subequations}
where given any Laurent series $L$ of $\lambda$, $(L)_+$ denotes its non-negative part
and the relation between $\hat M_k$ and $\mathcal A^{(n)}$ is explained in Section \ref{se:iso} below.

We interpret equation (\ref{i:comp-ii})  as flow on the dual space of the following twisted loop algebra:
\begin{eqnarray*}
{\frak g}_-&=&\left\{X(\lambda)=\sum_{-\infty}^{-1}X_i\lambda^i|X(\lambda)\sigma_1=
\sigma_1X(-\lambda)\right\}/{\frak g}_{2n+1},\qquad
\sigma_1=\left(\begin{array}{cc}0&1\\ 1&0\\\end{array}\right),\\
{\frak g}_{2n+1}&=&\left\{X(\lambda)=\sum_{-\infty}^{-2n-2}X_i\lambda^i|X(\lambda)\sigma_1=
\sigma_1X(-\lambda)\right\}
\end{eqnarray*}
Its Lie bracket and a precise description of the corresponding loop group can be found in Section
\ref{sec:coaj}. The dual space $\frak g_-^\ast$ of $\frak g_-$
can be identified with
\begin{eqnarray*}
{\frak g}_-^{\ast}&=&\left\{\Xi(\lambda)=\sum_{0}^{\infty}\Xi_i\lambda^i
\,| \Xi_i\in{\mathfrak{sl}}(2,\mathbb C),\,
\Xi(\lambda)\sigma_1=-\sigma_1\Xi(-\lambda)\right\}/{\frak g}_{2n+1}^{\ast},\\
{\frak g}_{2n+1}^{\ast}&=&\left\{\Xi(\lambda)=\sum_{2n+1}^{\infty}\Xi_i\lambda^i
\,| \Xi_i\in{\mathfrak{sl}}(2,\mathbb C),\,
\Xi(\lambda)\sigma_1=-\sigma_1\Xi(-\lambda)\right\}/{\frak g}_{2n+1}^{\ast},
\end{eqnarray*}
by the killing form in the loop algebra $\tilde{\mathfrak{sl}}(2,\mathbb C)$ (see (\ref{eq:pair}) below).

More precisely, to interpret  (\ref{i:comp-ii})  as flow on the $\frak{g}_-^{\ast}$, we denote by $A$ the dynamical part  of $\mathcal A^{(n)}$, i.e. $A=\left(\mathcal A^{(n)}\right)_+$, and we define
$$
\qquad B=
\left({{\mathcal A}^{(n)}\lambda^{1-2n}\over
4^{n}}\right)_-,
$$
where given any Laurent series $L$ of $\lambda$, $(L)_-$ denotes its  strictly negative part. In this way $A\in\frak g_-^\ast$ and $B\in \frak g_-$. Then we show that equation  (\ref{i:comp-ii})  is equivalent to
\begin{equation}\label{comp-iii}
\frac{{\rm d}A}{{\rm d}z}-\frac{\partial A}{\partial z}  =
[B,A]={\rm ad}^\ast_B A,
\end{equation}
so that the r.h.s. defines a vector field on the coadjoint orbit $\mathcal O_A$ of $A$ obtained by fixing the values of the parameters ${t}_1,\dots,{t}_{n-1}$ which are the Casimirs of the standard Poisson bracket on $\frak g_-^\ast$.  We then prove that the vector field
defined by ${\rm ad}^\ast_B A,$ is Hamiltonian with Hamiltonian given by
\begin{equation*}
H^{(n)}:= -\frac{1}{2\, 4^n} \tr\res\left(\lambda^{1-2n} \left(\A^{(n)}\right)^2\right),
\end{equation*}
where ${\rm Res}$ denotes the formal residue at $0$, i.e. the coefficient of the term in $\lambda^{-1}$.

Let us now describe our canonical coordinates
$P_1,\dots,P_n,Q_1,\dots,Q_n$ on the coadjoint orbit $\mathcal O_A$.
Our construction is based on the so--called algebro geometric
Darboux coordinates. The latter are the projections of the points in
the divisor of a line--bundle on the spectral curve
\cite{flM,nov-ve,adams}. However this classical construction in our
case is not straightforward: on one side there are too many points
in the divisor, on the other side the dependence of the matrix
entries of $\mathcal A^{(n)}$ on $w$ and its $z$--derivatives is
very complicated as it involves the recursive relation
(\ref{i:lenard}). In particular, this means that the matrix entries
are dependent on each other in a complicated way. We have resolved
both these problems by expressing the Lenard recursion operator $\mathcal L_n$ in terms of $\mathcal L_0,\dots,\mathcal L_{n-1}$ and their derivatives (see (\ref{eq:lzz}) below)  to
obtain the matrix entries as polynomials of the canonical
coordinates (see Theorem \ref{thm:ABB} below). It is worth noting that in \cite{FNR1} and \cite{FNR2}, a different set of canonical coordinates  on the coadjoint orbits of $\tilde{sl}(2,\mathbb{C})$ was found.  
Although these  preserve 
the Painlev\'e property, it is difficult to apply their construction to our case because of the complicated dependencies between the matrix entries of $\mathcal{A}^{(n)}$.

Resuming we prove that the map
\begin{equation}\label{i:map}
O_A\to (P_1,\dots,P_n,Q_1,\dots,Q_n)
\end{equation}
gives a system of rational Darboux coordinates on the coadjoint orbit. However this map depends on $z,t_1,\dots,t_{n-1}$ explicitly. This produces a shift $\delta H^{(n)}$ in the Hamiltonian $H^{(n)}$, so that the Hamiltonian
$\mathcal H^{(n)}$ appearing in (\ref{i:hpq}) is given by
$$
\mathcal H^{(n)}= H^{(n)}+ \delta H^{(n)}.
$$
Thanks to the fact that all formulae are explicit, we can compute this shift (see (\ref{hamPQ}) below).

The idea of interpreting the isomonodromic deformation equations as Hamiltonian flows on the coadjoint orbits of the loop group $\tilde{SL}(2,\mathbb C)$ was already used by Harnad and Routhier
\cite{har-rou} to study the Hamiltonian structure of the six classical Painlev\'e equations (see also \cite{har-d}). Later Krichever \cite{kri}
used the Lax representation approach to construct the isomonodromy equations for meromorphic connections with irregular and regular singularities on algebraic curves.

On the other hand, the coadjoint orbits of the dual loop algebras can also be thought of as moduli spaces of meromorphic connections on Riemann surfaces. Audin \cite{audin} generalized the Aityah-Bott symplectic structure on the moduli space of holomorphic connections to the case where the Riemann-surfaces can have boundaries. This symplectic structure was used by Hitchin to study the Schlesinger system \cite{hit}. Later, Boalch \cite{B} further generalized this symplectic structure to the moduli space of generic meromorphic connections. He then showed that these moduli spaces are isomorphic to the coadjoint orbits as symplectic manifolds, and that isomonodromic flows induce symplectomorphism between coadjoint orbits at different times. This abstractly indicates that one could express a general isomonodromic deformation as a time-dependent Hamiltonian flow on the dual loop algebra.
Woodhouse \cite{W} showed that the isomonodromic deformations with poles fixed are autonomous Hamiltonian systems w.r.t. the Konstant Kirillov Poisson structure on a central extension of the dual loop algebra (which is different from what we are considering in this paper).

Our construction is very explicit and allows us to go further to
interpret  the time--flows (\ref{i:laxMA}) as flows on the dual loop
algebra $\frak{g}_-^{\ast}$. This is not trivial because the term
$\p_{\lambda}\hat M^{(k)}$ is not tangent to the coadjoint orbit
$\mathcal O_A$.  To overcome this difficulty we introduce new
coordinates $u_1,\dots,u_{2n}$ on the coajoint orbit $\mathcal O_A$
and new times $s_1,\dots,s_{n-1}$ such that
$$
\p_{s_k}\mathcal{A}^{(n)}=[L_k,\mathcal{A}^{(n)}]+\p_{\lambda}L_k,\qquad
L_k= \left({\mathcal A}^{(n)}\lambda^{1-2k}\right)_+,\quad
\p_{s_k}^u\mathcal{A}^{(n)}=\p_{\lambda}L_k, \quad k=1,\ldots, n.
$$
where $\p_{s_k}^u$ denotes the partial derivative with respect to $s_k$ when $u$ is fixed.

We then interpret the equation
\begin{equation}\label{i:su}
\p_{s_k}\mathcal{A}^{(n)}-\p_{\lambda}L_k=(\p_{s_k}-\p_{s_k}^u)\mathcal{A}^{(n)}=[L_k,\mathcal{A}^{(n)}].
\end{equation}
as flow on the coadjoint orbit $\mathcal O_A$ and compute the corresponding Hamiltonians $h_1^{(n)},\dots,h_{n-1}^{(n)}$
$$
h^{(n)}_k =  \frac{1}{2} \tr\res\left(\lambda^{1-2k} \left(\A^{(n)}\right)^2\right),
\qquad k=1,\dots,n,
$$
so that in particular $H^{(n)}=-\frac{h^{(n)}_n}{4^n}$. Finally we show that the time--flows Hamiltonians
${\mathcal H}_1^{(n)},\dots,\mathcal H_{n-1}^{(n)}$ are given in terms of $h_1^{(n)},\dots,h_{n-1}^{(n)}$ (and their shifts due to the explicit dependence of (\ref{i:map}) on $z,t_1,\dots,t_{n-1}$) by a simple formula (see Corollary \ref{co:hamt} below).

This result gives an insight into how one could express a general isomonodromic deformation as a non--autonomous Hamiltonian system.

\begin{remark}
Note that $h_1^{(n)},\dots,h_{n}^{(n)}$ are spectral invariants. In the context of iso--spectral deformations, this can be used to show that the algebro--geometric Darboux coordinates  are separated for the isospectral system (see, for example \cite{skly1}). However, in the isomonodromic case, all the functions $h_1^{(n)},\dots,h_{n}^{(n)}$ are non--autonomous, i.e. they involve the variables $t_1,\dots, t_{n-1}$ and $z$ explicitly. Therefore we don't have separability in the sense of classical  mechanics. \end{remark}

This paper is organized as follows. In Section \ref{se:pii}, we
recall the derivation of the second Painlev\'e hierarchy as
self--similarity reduction of the mKdV hierarchy. In Section
\ref{se:iso}, we describe the monodromy problem associated to the
second Painlev\'e hierarchy. In Section \ref{sec:coaj}, we introduce
our  twisted loop algebra, interpret equation (\ref{i:comp-ii}) as
flow on its dual space $\mathfrak g^\ast_-$ study the Poisson
bracket on $\mathfrak g^\ast_-$. We prove that the parameters
$t_1,\dots, t_{n-1}$ belong to the kernel of such Poisson bracket
and characterize the symplectic leaves. In Section \ref{se:can}, we
introduce our coordinates $P_1,\dots,P_n, Q_1,\dots,Q_n$ and prove
that they are canonical with respect to the symplectic form on the
coadjoint orbit. In Section \ref{se:expl}, we obtain formula
(\ref{eq:lzz}) expressing the Lenard recursion operator $\mathcal
L_n$ in terms of $\mathcal L_0,\dots,\mathcal L_{n-1}$ and their
derivatives and we give the explicit formulae for the matrix entries
of $\mathcal A^{(n)}$ in terms of our canonical coordinates. We give
an explicit example to illustrate our procedure in detail. In
Section \ref{se:ham}, we compute the Hamiltonians $\mathcal
H^{(n)}$. In Section \ref{se:time}, we interpret equation
(\ref{i:su}) as  flow on the coadjoint orbit, compute the
corresponding Hamiltonians $h_1^{(n)},\dots,h_{n-1}^{(n)}$ and show
that they are spectral invariants. Finally we obtain the
Hamiltonians $\mathcal H^{(n)}_1,\dots, \mathcal H^{(n)}_{n-1}$.  We
follow all details of our construction in an explicit example.

\vskip 0.2 cm
\noindent{\bf Acknowledgments.} The authors are grateful to H.  Flaschka,  M. Talon, J. Harnad,  A. Hone and to V. Kuznetsov
for helpful conversations. This research was sponsored by the ESF grant MISGAM, the Marie Curie network ENIGMA and by EPSRC.

\section{The second $\Pt$ hierarchy}\label{se:pii}
The second Painlev\'e hierarchy is obtained as {\it
self-similarity reduction}\/ of the modified Korteweg-de Vries
(mKdV) hierarchy  (see
\cite{refAS77,refAirault,FN}):
\begin{equation}
\pd{T_{n+1}} v +\pd{x}\left(\pd{x}+2v\right)\mathcal{R}_n
\left[v_x-v^2\right]=0,
   \qquad{}
n=0,1,2,3,\ldots \label{mKdVh}
\end{equation}
where $\mathcal{R}_n$ satisfies the Lenard recursion relation
\cite{refLax}
\begin{equation}\label{lenardkdv}
\pd{x}\mathcal{R}_{n+1} =\left(\pderiv{^3}{x^3}+4(v_x-v^2)\pd{x}+
2(v_x-v^2)_x\right)\mathcal{R}_n,\qquad
\mathcal{R}_0[u] = {\textstyle\frac12}.
\end{equation}
Each equation in the mKdV hierarchy defines a Hamiltonian flow and
can be viewed as a symmetry for all others. Consider the space of stationary
solutions w.r.t. the symmetry defined by the n-th mKdV equation, i.e. the space of solutions
$v(x;T_1,T_2,\dots)$ such that $\frac{\partial v}{\partial T_{n+1}}  =0$.
Due to the fact that all Hamiltonian flows
commute,  all other elements of the hierarchy  can be restricted to this space.

There are also other symmetries acting on the mKdV hierarchy. They are called
Virasoro symmetries. The $n$-th Virasoro symmetry is given by the following
infinitesimal generator
$$
\frac{\rm d}{{\rm d}s_n}:=\sum_{l=0}^n(2 l+1) T_{l+1} \frac{\partial}{\partial T_{l+1}}.
$$
The stationary solutions w.r.t. this generator are by definition
such that $\frac{{\rm d}v}{{\rm d}s_n}\equiv 0$. They satisfy
$$
\frac{{\rm d}v}{{\rm d}s_n}=-\sum_{l=0}^n(2 l+1)
T_{l+1}\pd{x}\left(\pd{x}+2v\right)\mathcal{R}_l\left[v_x-v^2\right]=0,
$$
and after integration
\begin{equation}\label{protopii}
 -\sum_{l=0}^n(2 l+1)
T_{l+1}\left(\pd{x}+2v\right)\mathcal{R}_l\left[v_x-v^2\right]=\alpha_n,
\end{equation}
where $\alpha_n$ is some constant.\footnote{Recently S. Kakei
\cite{kakei}, proposed a  new way to obtain the second Painlev\'e
equation directly as a reduction of mKdV, without integration. In
his difference--operator formulation, the constant $\alpha_1$
appears as a parameter in the symmetry reduction, and not as
integration constant. It would be interesting to see whether this
construction can be used to produce the whole $\Pt$ hierarchy.} From
the $n=0$ equation of the mKdV hierarchy, we can set $T_1=-x$  so
that (\ref{protopii}) is an ODE in the variable $x$ depending on
some extra parameters $T_2,\dots,T_{n+1}$. The parameter $T_{n+1}$
can be absorbed by the following symmetry reduction (see
\cite{refCJP} for details):
\begin{subequations}\label{self-sim}
\begin{eqnarray}
&&
v(x,T_{n+1})={w(z)\over\left[(2n+1)T_{n+1}\right]^{1/(2n+1)}},\qquad
z={x\over\left[(2n+1)T_{n+1}\right]^{1/(2n+1)}},\\
&&
{\mathcal R}_l\left[v_x-v^2\right]= \frac{1}{\left[(2n+1)T_{n+1}\right]^{\ifrac{2l}{2n+1}}}\,
      {\mathcal L}_l[w_z - w^2],\\
      &&
{t}_0=-z,\qquad
{t}_{l}:=\frac{ (2 l+1) T_{l+1}}
{\left[(2n+1)T_{n+1}\right]^{\ifrac{(2l+1)}{2n+1}}},\quad  l=1,\dots,n,\quad {t}_n=1.
\end{eqnarray}
\end{subequations}
In this way we obtain the Second Painleve Hierarchy:\footnote{To obtain exact
solutions of the $n$-th mKdV equation, one fixes the values of ${t}_1,\dots, {t}_{n-1}$. As a consequence, often in the literature the second Painlev\'e hierarchy is presented with
 ${t}_1=\dots={t}_{n-1}=0$. We will leave ${t}_1,\dots, {t}_{n-1}$ free to vary instead.}
\begin{equation}\label{timepii}
 \Ptn :\quad\quad
\left(\od{z}+2 w\right) \Lr_{n}\left[w_{z}-w^{2}\right]
+\sum_{l=1}^{n-1} {t}_{l}\left(\od{z}+2 w\right)
 \Lr_{l}\left[w_{z}-w^{2}\right]
=z w+\alpha_{n},\qquad n\ge 1,
\end{equation}
where $\alpha_{n}$ are constants and $\Lr_{n}$ is the
operator defined by
\begin{equation}\label{lenard}
\frac{{\rm d}}{{\rm d}z}
{\mathcal L}_{n+1}= \left(\frac{{\rm d^3}}{{\rm d}z^3}+4(w_z-w^2)
\frac{{\rm d}}{{\rm d}z}+2(w_z-w^2)_z\right){{\mathcal L}}_n,
\qquad {\mathcal L}_0[w_z-w^2] = {\textstyle\frac12}.
\end{equation}

\begin{example}
For $n=1$, equation (\ref{timepii}) is $\Pt$:
$$
w_{zz}-2 w^3 = z w+\alpha_1.
$$
For $n=2$, it is:
$$
{t}_1 (w_{zz}-2 w^3) +  (w_{zzzz} -10 w w_z^2-10 w^2 w_{zz}
+6 w^5) = z w +\alpha_2.
$$
\end{example}

In the case of the Virasoro symmetries, only the first $n$ flows
of the mKdV hierarchy can be restricted to the space of such
stationary solutions. These give the flows in ${t}_1,\dots,{t}_{n-1}$:
\begin{equation}\label{timeflows}
(2k+1)\frac{\p w}{\p {t}_k} + \p_z\left(\p_{z}+2 w\right) \Lr_{k}\left[w_{z}-w^{2}\right]=0,\qquad k=1,\dots,n-1.
\end{equation}

We shall call the flows in ${t}_1,\dots,{t}_{n-1}$ {\it
time-flows}.

\begin{remark}
Another Painlev\'e hierarchy containing $\Pt$ as its first element has been introduced by Gordoa,
Joshi and Pickering \cite{GJP} by generalizing the isomonodromic deformation equations by Jimbo Miwa. However it is not clear if their hierarchy is different or not from the one studied in this paper. We know that Koike from Kyoto University is building the Hamiltonian structure of Gordoa,
Joshi and Pickering hierarchy as confluence procedure of the Garnier systems. Once he is successful, we may try to see whether our Hamiltonian system and Koike's are related by a canonical transformation. If not, the question of whether the $\Pt$ hierarchy considered in this paper may or may not arise as confluence limit of the Garnier system remains open.
\end{remark}


\section{Isomonodromic Problem for the $\Pt$ Hierarchy}\label{se:iso}

The isomonodromic deformation problem for the second Painlev\'e hierarchy with
${t}_1=\dots={t}_{n-1}=0$ was derived in \cite{refCJM} and in \cite{Kud} following
the approach proposed in \cite{AKNS} starting form the isomonodromy deformation problem given in \cite{FN} for the second Painlev\'e equation. Here we generalize the
construction of  \cite{refCJM}  to the case of generic values ${t}_1,\dots, {t}_{n-1}$ (for details see the Appendix \ref{app-a}).

The isomonodromic deformation problem for the $\Pt$ Hierarchy is
the following:
\begin{subequations}\label{isomonodromy}
\begin{eqnarray}
{\partial\Psi\over\partial z}&=&{\mathcal{B}}\Psi=\left(\begin{array}{cc}
         -\lambda& w\\
         w&\lambda
       \end{array}
       \right)\Psi,\label{isomonodromy.1}\\
{\partial\Psi\over\partial\lambda}&=&{\mathcal{A}}^{(n)}\Psi
= \frac{1}{\lambda}\left[
\left(\begin{array}{cc}
      -\lambda z&-\alpha_n\\
         -\alpha_n&\lambda z
       \end{array}
       \right) + M^{(n)}+\sum_{l=1}^{n-1} {t}_{l} M^{(l)}
\right]\Psi,
\label{isomonodromy.2}\\
(2k+1)\frac{\p\Psi}{\p {t}_k}&=& \left(M^{(k)} -\left(
\begin{array}{cc} 0&(\partial_z+2 w){\mathcal L}_k\\
(\partial_z+2 w){\mathcal L}_k&0
\end{array}\right) \right)\Psi,
\label{isomonodromy.3}
\end{eqnarray}
\end{subequations}
where
$$
M^{(l)}= \left(\begin{array}{cc}
       \sum_{j=1}^{2l+1}{A}_{j}^{(l)}\lambda^{j}&
\sum_{j=1}^{2l}{B}_{j}^{(l)}\lambda^{j}\\
       \sum_{j=1}^{2l}{C}_{j}^{(l)}\lambda^{j}
&-\sum_{j=1}^{2l+1}{A}_{j}^{(l)}\lambda^{j}
\end{array}\right),
$$
with
\begin{subequations}\label{isomonodromy.4}
\begin{eqnarray}
 {A}_{2 l+1}^{(l)}&\!\!\!=\!\!\!& 4^l, \qquad  {A}_{2 k}=0,
\quad\forall\, k=0,\dots,l,\\
 {A}_{2 k+1}^{(l)}&\!\!\!=\!\!\!&\frac{4^{k+1}}{2}
\left\{ {\mathcal L}_{l-k}\left[w_z-w^2\right]-\od{z}
\left(\od{z}+2 w\right)
{\mathcal L}_{l-k-1}\left[w_z-w^2\right]\right\},
\quad k=0,\dots,l-1,\\
 {B}_{2 k+1}^{(l)}&\!\!\!=\!\!\!& \frac{4^{k+1}}{2}\od{z}\left(\od{z}+2
w\right) {\mathcal L}_{l-k-1}\left[w_z-w^2\right],
\quad k=0,\dots,l-1,\\
 {B}_{2 k}^{(l)}&\!\!\!=\!\!\!&-4^k\left(\od{z}+2
w\right) {\mathcal L}_{l-k}\left[w_z-w^2\right],
\quad k=1,\dots,l,\\
 {C}_{2 k+1}^{(l)}&\!\!\!=\!\!\!&- {B}_{2 k+1},
\quad k=0,\dots,l-1,\\
 {C}_{2 k}^{(l)}&\!\!\!=\!\!\!&  {B}_{2 k},
\quad k=0,\dots,l.
\end{eqnarray}
\end{subequations}

The compatibility between (\ref{isomonodromy.1}) and (\ref{isomonodromy.2}) gives
\begin{equation}\label{comp-ii}
\frac{\partial{\mathcal A}^{(n)}}{\partial z}  -\frac{\partial{\mathcal B}}{\partial \lambda}=
[{\mathcal B},{\mathcal A}^{(n)}],
\end{equation}
which gives (\ref{timepii}) (see \cite{refCJM}).  Equation
(\ref{timeflows}) is obtained as compatibility between
(\ref{isomonodromy.1}) and (\ref{isomonodromy.3}):
\begin{equation}\label{eq-MB}
(1+2k)\p_{{t}_k}\mathcal{B}-\p_z \hat M^{(k)}=[\mathcal{B},\hat
M^{(k)}],
\end{equation}
where for brevity we put
$$
\hat M_k=M^{(k)} - \left(
\begin{array}{cc} 0&(\partial_z+2 w){\mathcal L}_k\\
(\partial_z+2 w){\mathcal L}_k&0
\end{array}\right).
$$
Finally the compatibility between (\ref{isomonodromy.2}) and
(\ref{isomonodromy.3}) is
\begin{equation}\label{laxMA}
\p_{\lambda}\hat M^{(k)}-(2k+1)\p_{{t}_k}\mathcal{A}^{(n)}=-[\hat M^{(k)},\mathcal{A}^{(n)}].
\end{equation}

The proof of the fact that equations (\ref{comp-ii}), (\ref{eq-MB}) and (\ref{laxMA}) are indeed consistent is sketched in the Appendix \ref{app-a}.

To simplify our computations, it is convenient to introduce some new notations. We define
\begin{eqnarray}
&&
a^{(n)}_{2 k+1}=\sum_{l=1}^{n} {t}_l A_{2k+1}^{(l)},\qquad k=1,\dots,n,\qquad
a^{(n)}_{1}=\sum_{l=1}^{n} {t}_l A_{1}^{(l)}-z,\nn\\
&&
b^{(n)}_{2 k+1}=\sum_{l=1}^{n} {t}_l B_{2k+1}^{(l)},\qquad k=0,\dots,n-1,\nn\\
&&
b^{(n)}_{2 k}=\sum_{l=1}^{n} {t}_l B_{2k}^{(l)},\qquad k=1,\dots,n,\qquad
b^{(n)}_{0}=-\alpha_n,\nn
\end{eqnarray}
where ${t}_n=1$, so that we can write
\begin{equation}\label{matA}
{\mathcal A}^{(n)}:=\left(\begin{array}{cc}
\sum_{k=0}^n a^{(n)}_{2 k+1}\lambda^{2k}&\sum_{k=0}^n b^{(n)}_{2 k}\lambda^{2k-1}+
\sum_{k=0}^{n-1} b^{(n)}_{2 k+1}\lambda^{2k}\\
\sum_{k=0}^n b^{(n)}_{2 k}\lambda^{2k-1}-
\sum_{k=0}^{n-1} b^{(n)}_{2 k+1}\lambda^{2k}&
-\sum_{k=0}^n a^{(n)}_{2 k+1}\lambda^{2k}\\
\end{array}\right).
\end{equation}

\section{Coadjoint orbit interpretation}\label{sec:coaj}

In this section we show that the Hamiltonian structure of the second Painlev\'e hierarchy can be derived from the one on an appropriate dual loop algebra.

Since our matrices ${\mathcal A}^{(n)}$ depend on $z$ and $\lambda$ and the variable $z$ appears both implicitly, through $w(z)$ and its derivatives, and explicitly, we need to introduce some notation.
Given any function $f$ of $\lambda,z,w,w_z,\dots$,
let us denote the partial derivative of $f$ w.r.t. $z$ as follows:
$$
\partial_z f:=\frac{\partial f}{\partial z}+\frac{\partial f}{\partial w}w_z+
\frac{\partial f}{\partial w_z}w_{zz}+\dots,
$$
and  $\partial_z^w f$ the partial derivative of $f$ considered as a differential polynomial of $w$ depending on $z,\lambda$:
$$
\partial_z^w f:=\frac{\partial f}{\partial z}.
$$
Analogously $\partial_\lambda f$ denotes the partial derivative of $f$ w.r.t. $\lambda$.
Then given the matrices $\mathcal B$ and ${\mathcal A}^{(n)}$ as in (\ref{isomonodromy}) and (\ref{isomonodromy.4}), one has
\begin{equation}\label{crucial}
\partial_z^w {\mathcal A}^{(n)}=\partial_\lambda{\mathcal B},
\end{equation}
so that equation ({\ref{comp-ii}) is equivalent to
\begin{equation}\label{lax}
(\partial_z-\partial_z^w) {\mathcal A}^{(n)}=[{\mathcal B},
{\mathcal A}^{(n)}].
\end{equation}

\begin{remark}
This phenomenon, i.e. equation (\ref{crucial}), is a common feature of all Painlev\'e equations and it was used  in  \cite{har-rou} to find the algebro--geometric Darboux coordinates for the six Painlev\'e equations. As far as we know, a proof of the fact that  (\ref{crucial}) is a common feature of all the
isomonodromic deformations equations is still missing.
\end{remark}

We are now going to interpret the evolution along $(\partial_z-\partial_z^w) $ as a vector field on a coadjoint orbit of an element of an appropriate twisted loop algebra. Let $LG$ be the group of smooth maps $f$ from $S^1$ to $SL_2$ such that
$$
f(\lambda)\sigma_1(f(-\lambda))^{-1}=I,\qquad \sigma_1=\left(\begin{array}{cc}0&1\\ 1&0\\\end{array}\right),
$$
and $\lambda$ is considered as a parameter on $S^1$. Denote by
$L_{2n+2}G$ the subgroup of maps of the form
$f=I+\lambda^{-2n-2}f_{\infty}$, where $f_{\infty}$ is holomorphic
outside $S^1$ and let $\frak{g}_{2n+2}$ be its Lie algebra:
$$
{\frak g}_{2n+2}=\left\{X(\lambda)=
\sum_{-\infty}^{-2n-2}X_i\lambda^i| X_i\in{\mathfrak{sl}}(2,\mathbb
C),\, X(\lambda)\sigma_1=\sigma_1X(-\lambda)\right\}.
$$
Then let
$G$ be the quotient of these 2 groups
$$
G=LG/L_{2n+2}G.
$$
Its Lie algebra is given by
$$
{\frak g}=\left\{X(\lambda)=
\sum_{-\infty}^{\infty}X_i\lambda^i| X_i\in{\mathfrak{sl}}(2,\mathbb C),\,
X(\lambda)\sigma_1=\sigma_1X(-\lambda)\right\}/{\frak g}_{2n+2},
$$
with Lie bracket defined as:
$$
[X(\lambda),\tilde X(\lambda)]= \sum_{i=-2n-1}^{\infty}\left(
\sum_{k=-2n-1}^{i+2n+1} [X_{k},\tilde
X_{i-k}]\right)\lambda^i\mod{\frak g}_{2n+2},
$$
which obviously gives $[X(\lambda),\tilde X(\lambda)]\in\mathfrak g$ and satisfies the Jacobi identity.
The dual space
$\mathfrak g^\ast$ can be identified with
\begin{eqnarray*}
{\frak g}^\ast&=&\left\{\Xi(\lambda)=
\sum_{-\infty}^{\infty}\Xi_i\lambda^i\,|\, N\in{\mathbb N},\,
\Xi_i\in{\mathfrak{sl}}(2,\mathbb C),\,
\Xi(\lambda)\sigma_1=-\sigma_1\Xi(-\lambda)\right\}/\frak{g}_{2n+2}^{\ast},\\
{\frak g}_{2n+2}^{\ast}&=&\left\{X(\lambda)=
\sum_{2n+1}^{\infty}X_i\lambda^i| X_i\in{\mathfrak{sl}}(2,\mathbb
C),\, X(\lambda)\sigma_1=\sigma_1X(-\lambda)\right\}.
\end{eqnarray*}
by the following pairing
\begin{equation}\label{eq:pair}
\langle X(\lambda),\Xi(\lambda)\rangle := {\rm Tr}\left({\res}X(\lambda)\Xi(\lambda)\right),
\quad\forall\, X(\lambda)\in\mathfrak g,\,  \Xi(\lambda)\in\mathfrak g^\ast,
\end{equation}
where $\res$ indicates the formal residue, i.e. the coefficient of the $\lambda^{-1}$ term.
Consider the subalgebra
\begin{equation}\label{lie2}
{\frak
g}_-=\left\{X(\lambda)=\sum_{-\infty}^{-1}X_i\lambda^i|X(\lambda)\sigma_1=\sigma_1X(-\lambda)\right\}/{\frak
g}_{2n+2},
\end{equation}
its dual space can be identified with
\begin{equation}\label{lie3}
{\frak
g}_-^{\ast}=\left\{\Xi(\lambda)=\sum_{0}^{\infty}\Xi_i\lambda^i
|\Xi_i\in{\mathfrak{sl}}(2,\mathbb C),\,
\Xi(\lambda)\sigma_1=-\sigma_1\Xi(-\lambda)\right\}/{\frak
g}_{2n+2}^{\ast}.
\end{equation}
An element $X$ in the Lie algebra ${\frak g}$ acts on an element
$\Xi\in {\frak g}^{\ast}$ by the coadjoint action
\begin{eqnarray}\label{eq:coadjoint}
\left<{\rm ad}^{\ast}_X\Xi,Y\right>:=-\left<\Xi,[X,Y]\right>=\left<[X,\Xi],Y\right>
\end{eqnarray}
for any $Y\in{\frak g}$. This shows that for every $X\in{\frak g}$, $\Xi\in\frak g^\ast$
\begin{eqnarray*}
[X,\Xi]={\rm ad}^\ast_X \Xi \in{\mathfrak g}^\ast.
\end{eqnarray*}
When we restrict the coadjoint action to the subalgebra ${\frak g}_-$
and to its dual space ${\frak g}_-^{\ast}$, we obtain the following identification
\begin{eqnarray}\label{eq:tangent}
[X_-,\Xi]_+={\rm ad}^\ast_{X_-} \Xi ,\quad \Xi\in {\frak g}_-^{\ast},
\quad X_-\in{\frak g}_-
\end{eqnarray}
where $(\cdot)_+$ is the projection from ${\frak g}^{\ast}$ onto
${\frak g}_-^{\ast}$ and  $(\cdot)_-$ denotes the projection onto ${\frak g}_-$.

\begin{lemma}
Given the matrices $\mathcal B$ and ${\mathcal A}^{(n)}$ as in (\ref{isomonodromy}) and (\ref{isomonodromy.4}), one has
\begin{equation}\label{eq:coadjtang}
[{\mathcal B},{\mathcal A}^{(n)}]= {\rm ad}^\ast_B A,
\end{equation}
where $B=\left({{\mathcal A}^{(n)}\lambda^{-2n+1}\over 4^{n}}\right)_-\in{\mathfrak g}_-$ and
$A=\left({\mathcal A}^{(n)}\right)_+\in{\mathfrak g}^\ast_-$, which is the dynamical part of ${\mathcal
A}^{(n)}$.
\end{lemma}

\begin{proof} Using  (\ref{isomonodromy}) and (\ref{isomonodromy.4}), we notice that:
$$
{\mathcal B}=-\left({{\mathcal A}^{(n)}\lambda^{-2n+1}\over
4^{n}}\right)_+.
$$
Then using the Drinfeld--Sokolov trick:
\begin{eqnarray}\label{eq:coadjtang1}
[{\mathcal B},{\mathcal A}^{(n)}]&=&-\left[\left({{\mathcal A}^{(n)}\lambda^{-2n+1}\over
4^{n}}\right)_+,{\mathcal A}^{(n)}\right]=\nonumber\\
&=&\left[\left({{\mathcal A}^{(n)}\lambda^{-2n+1}\over 4^{n}}\right)_-,{\mathcal
A}^{(n)}\right]=\nonumber\\
&=&\left[\left({{\mathcal A}^{(n)}\lambda^{-2n+1}\over 4^{n}}\right)_-,\left({\mathcal
A}^{(n)}\right)_+\right],
\end{eqnarray}
where the last step is due to the fact that $\left({{\mathcal A}^{(n)}\lambda^{-2n+1}\over 4^{n}}\right)_-$
commutes with $\left({\mathcal A}^{(n)}\right)_-$.
\end{proof}

\begin{remark}
The fact that we can neglect the singular part of $\mathcal A^{(n)}$ at $\lambda=0$ is more general than in the above proof. Suppose we want to compute ${\rm ad}_{X_k}^\ast \mathcal A^{(n)}$ for
$X_k=\left({{\mathcal A}^{(n)}_t\lambda^{-2k+1}\over 4^{k}}\right)_-$,
then for every $Y\in\mathfrak g_-$ we have
\begin{eqnarray*}
\langle{\rm ad} _{X_k}^\ast \mathcal A^{(n)},Y\rangle&=&
\left<\left[\left({{\mathcal A}^{(n)}\lambda^{-2k+1}\over
4^{k}}\right)_-,{\mathcal A}^{(n)}\right],Y\right>\\
&=& \left<\left[\left({{\mathcal A}^{(n)}\lambda^{-2k+1}\over
4^{k}}\right)_-,\left({\mathcal A}^{(n)}\right)_+\right],Y\right>,
\end{eqnarray*}
because the singular part of  $\mathcal A^{(n)}$
does not contribute to the residue.
\end{remark}

Similarly it is easy to see that
$$
\left(\partial_z-\partial_z^w\right){\mathcal A}^{(n)}=\left(\partial_z-\partial_z^w\right) A.
$$
Resuming, we proved the following Lemma:

\begin{lemma}\label{lm:coaj}
The monodromy preserving deformation equation (\ref{comp-ii}) is the same as
\begin{equation}\label{lax2}
\left(\partial_z-\partial_z^w\right) A={\rm ad}^\ast_B A,
\end{equation}
where $A=\left(\mathcal A^{(n)}\right)_+\in{\mathfrak g}_-^\ast$ is the dynamical part of
$\mathcal A^{(n)}$, and
$B=\left(\frac{\mathcal A^{(n)}\lambda^{1-2n}}{4^n}\right)_-\in{\mathfrak g}_-$.
\end{lemma}

This Lemma allows us to interpret the evolution along $(\partial_z-\partial_z^w) $ as a vector field on a coadjoint orbit of the twisted loop algebra $\mathfrak g_-$.

Let us now recall the Poisson structure on ${\frak g}_-^{\ast}$.
This is fairly standard (see for example the beautiful book
\cite{BBT}), but we recall some details here in order to fix
notations and adapt the computations to our special case.

The Poisson structure on $\frak{g}_-^{\ast}$ is given by observing
that every $X\in\mathfrak g_-$ defines a linear function $X_\ast$ on
$\mathfrak g_-^\ast\ni\Xi$:
$$
X_\ast:\begin{array}{lcl}
\mathfrak g_-^\ast&\to&{\mathbb C}\\
\Xi&\to&\langle \Xi,X\rangle.\\
\end{array}
$$
This fact allows one to identify $\mathfrak g_-^{\ast^\ast}$ with $\mathfrak g_-$ and to define the Poisson bracket between two linear functions on $\mathfrak g_-^\ast$ as S. Lie did
$$
\{X_\ast, Y_\ast\}(\Xi):=\left<\Xi,[X,Y]\right>.
$$
The Poisson bracket between two functions $f$ and $g$ on $\mathfrak g_-^\ast$ is given by
\begin{eqnarray}\label{eq:poisson}
\left\{f,g\right\}(\Xi)=\left<\Xi,[{\rm d}f,{\rm d} g]\right>
\end{eqnarray}
where the differential ${\rm d}f$ of a function $f$ on $\frak g_-^\ast$ is a linear function
${\rm d}f\in{\mathfrak g_-^\ast}^\ast\sim\mathfrak g_-$ defined by
\begin{equation}\label{diff}
\langle {\rm d f},\delta \Xi\rangle:=f\left(\Xi+\delta_X \Xi\right)-f(\Xi)+\mathcal O(\delta_X \Xi)^2,
\end{equation}
where
$$
\delta_X \Xi:= {\rm ad}^\ast_X \Xi \in{\mathfrak g}_-^\ast.
$$
In particular one has ${\rm d} X_\ast=X$.

It is well known that this Poisson bracket is degenerate and its symplectic leaves are the coadjoint orbits of its elements. The kernel of this bracket
consisting of the {\it Casimirs,}\/ i.e.  functions $f$ such  that
$$
{\rm ad}^\ast_X \Xi ({\rm d}f)=0,\qquad \forall X\in\frak g_-,\, \Xi\in \frak g_-^\ast.
$$

\begin{lemma}\label{le:casimir}
The times ${t}_1,\dots,{t}_{n-1}$ are the Casimirs of the Poisson bracket (\ref{eq:poisson}).
\end{lemma}

\begin{proof}
We show that the Hamiltonian vector fields generated by ${t}_1,\dots,{t}_{n-1}$ are zero.

Denote the eigenvalues of ${\mathcal A}(\lambda)$ by
$\pm\mu(\lambda)$. The polynomial part of $\mu(\lambda)$ is a
polynomial of order $2n$ and all the coefficients of the odd
positive powers of $\lambda$ are zero. Therefore the polynomial part
of $\mu(\lambda)$ has exactly only $n+1$ non zero coefficients, of
which the first one is $4^n$. Our claim is that all the other
coefficients of the positive powers of $\lambda$ give our times
${t}_1,\dots,{t}_{n-1}$ (we shall see in Section \ref{se:ham}
that the coefficient of the $-2$ power of  $\lambda$ is the
Hamiltonian in the variable $z$).

In fact, as proved in corollary \ref{co:si} below, the times ${t}_i$ are given by the `spectral residue formula' \cite{har} as follows.
\begin{eqnarray*}
{t}_l=\frac{1}{4^l}\res_{\infty}\lambda^{-2l-1}\mu(\lambda) d\lambda=
\langle A, \frac{\lambda^{-2l-1}}{2\,4^l}\Psi\sigma_3\Psi^{-1},
\rangle
\end{eqnarray*}
where $\Psi$ is the eigenvector matrix of ${\mathcal A}$ and $\sigma_3=\left(\begin{array}{cc}
1&0\\0&-1\\\end{array}\right)$.
Since $ \frac{\lambda^{-2l-1}}{2\, 4^l}\Psi\sigma_3\Psi^{-1}$ commutes with ${\mathcal A}(\lambda)$, it produces a trivial vector field and therefore ${t}_l$ is a Casimir.
\end{proof}

Thanks to the above Lemma, the coadjoint orbits are obtained by fixing the values of
${t}_1,\dots,{t}_{n-1}$:
$$
{\mathcal O}_A:= \left\{
\left(\begin{array}{cc}
\sum_{k=0}^n a^{(n)}_{2 k+1}\lambda^{2k}&\sum_{k=1}^n b^{(n)}_{2 k}\lambda^{2k-1}+
\sum_{k=0}^{n-1} b^{(n)}_{2 k+1}\lambda^{2k}\\
\sum_{k=1}^n b^{(n)}_{2 k}\lambda^{2k-1}-
\sum_{k=0}^{n-1} b^{(n)}_{2 k+1}\lambda^{2k}&
-\sum_{k=0}^n a^{(n)}_{2 k+1}\lambda^{2k}\\
\end{array}\right)\bigg|_{{t}_1={t}_1^0,\dots,{t}_{n-1}={t}_{n-1}^0}
\right\}
$$
The dimension of the coadjoint orbits is $2n$. It is well known that
the  Poisson bracket (\ref{eq:poisson}) restricted to the coadjoint
orbits is non-degenerate, so that it defines the so-called
Kostant-Kirillov symplectic structure $\omega$ on them. We will
write
$$
\omega(f,g) = \{f,g\},
$$
for every pair of functions on the coadjoint orbit.

To compute the Poisson brackets between the coefficients
$a_{2k+1}^{(n)}$, $b_{2k}^{(n)}$ and $b_{2k+1}^{(n)}$, we observe
that their differentials are
\begin{eqnarray}\label{eq:grad}
{\rm d} a_{2k+1}^{(n)}&=&{1\over 2}(E_{11}-E_{22})\lambda^{-(2k+1)}
\qquad\hbox{for}\quad 0\leq k\leq n, \nonumber\\
{\rm d} b_{2k+1}^{(n)}&=&{1\over 2}(-E_{12}+E_{21})\lambda^{-(2k+1)}
\qquad\hbox{for}\quad 0\leq k\leq n-1, \\
{\rm d} b_{2k}^{(n)}&=&{1\over 2}(E_{12}+E_{21})\lambda^{-2k}
\qquad\hbox{for}\quad 0\leq k\leq n, \nonumber
\end{eqnarray}
where with a slight abuse of notation we are calling  $a_{2k+1}^{(n)}$,
$b_{2k}^{(n)}$ and $b_{2k+1}^{(n)}$ the elements of ${\mathfrak g_-^\ast}^\ast\sim\mathfrak g_-$
which applied to $\Xi$ produce the coefficients  $a_{2k+1}^{(n)}$,
$b_{2k}^{(n)}$ and $b_{2k+1}^{(n)}$ respectively.

By using these gradients, we can compute the Poisson brackets
between the matrix entries
\begin{eqnarray}\label{eq:bracket}
\left\{a_{2k+1}^{(n)},b_{2l+1}^{(n)}\right\}&=&-b^{(n)}_{2(k+l+1)},
\qquad\hbox{for}\quad 0\leq k\leq n, \quad 0\leq l\leq n-1, \quad k+l\leq n-1, \nonumber\\
\left\{a_{2k+1}^{(n)},b_{2l}^{(n)}\right\}&=&-b^{(n)}_{2(k+l)+1}
\qquad\hbox{for}\quad 0\leq k\leq n,\quad 1\leq l\leq n, \quad k+l\leq n-1, \\
\left\{b_{2k}^{(n)},b_{2l+1}^{(n)}\right\}&=&a^{(n)}_{2(k+l)+1}
\qquad\hbox{for}\quad 1\leq k\leq n,\quad 0\leq l\leq n-1, \quad k+l\leq n, \nonumber
\end{eqnarray}
while all the other brackets vanish.

\section{Canonical coordinates for the isomonodromic deformations}\label{se:can}

Our first attempt to build the canonical coordinates for the second Painlev\'e hierarchy is to use the general framework of the algebro--geometric Darboux coordinates (see \cite{flM,nov-ve,adams}). In this setting one considers the spectral curve
\begin{equation}\label{eq:spectralcurve}
\Gamma(\mu,\lambda)=\left\{
\det(\mu-\mathcal{A}^{(n)}(\lambda))=0
\right\}=\left\{\mu^2= -\det\left({\mathcal A}^{(n)}(\lambda)\right)\right\}.
\end{equation}
The {\it characteristic equation}
$\mu^2 = -\det\left({\mathcal A}^{(n)}(\lambda)\right)$ defines the eigenvalue
$\mu(\lambda)$ of $A(\lambda)$ as a
function on the corresponding $2$-sheeted Riemannian surface of genus $g$.
The {\it Baker--Akhiezer function}
$\psi(\lambda)$ is defined then as the eigenvector of
$\mathcal{A}^{(n)}(\lambda)$
$$
\mathcal{A}^{(n)}(\lambda) \psi(\lambda)=\mu(\lambda)\psi(\lambda)
$$
corresponding to the eigenvalue $\mu(\lambda)$. Generally, $\psi$ has $g+1$ poles.

Following  \cite{skly1}, let us briefly illustrate how to construct
canonical coordinates $p_1,\dots,p_g$ and $q_1,\dots,q_g$ on the cotangent bundle of the Jacobian $T^{\ast}J$ of the curve $\Gamma$.

Denote by $q$ the $\lambda$--projection of the generic point in the divisor of $\psi$. We fix the following normalization
$$
(c_1, c_2 )\cdot \psi(q)=1,
$$
for some choice of $c_1,c_2$. The
$q_j$ variables are the roots of
\begin{equation}\label{sk:q}
c_1^2 A_{12}(q_j)-c_1 c_2 (A_{11}(q_j)-A_{22}(q_j))
         - c_2^2 A_{21}(q_j)=0,
\end{equation}
while the $p_j$ variables are the eigenvalues
\begin{equation}\label{sk:p}
     p_j=\left(A_{11}(q_j)-\frac{c_1}{c_2}A_{12}(q_j)\right).
\end{equation}
Choosing
the normalization $c_1= -c_2=1$ we get roots $q_1,\dots,q_{2n}$ such that $q_{n+j}=-q_j$, $j=1,\dots,n$. They are the roots of the following equation:
\begin{equation}\label{eq:smallq}
 \sum_{k=0}^{n-1}(b^{(n)}_{2k+1}+a^{(n)}_{2k+1}) \lambda^{2k}+
a^{(n)}_{2n+1}\lambda^{2n}=0.
\end{equation}
The corresponding $p_j$ are given by
$$
p_j=\sum_{k=0}^{n} b^{(n)}_{2k} q_j^{2k-1}.
$$

In the generic case, it is well-known that the coordinates $q_1,\dots,q_{2n}$, $p_1,\dots,p_{2n}$ are canonical with respect to the Konstant-Kirillov Poisson structure as well
\begin{equation}\label{sk:poisson}
\{p_i,p_j\}=\{q_i,q_j\}=0,\quad \{p_i,q_j\}=\delta_{ij},
\end{equation}
however, a proof of this fact  for non--generic cases is still missing.

Generically, the dimension $2g$ of  $T^\ast J$ coincides with the dimension of the symplectic leaves in the coadjoint orbit associated to (\ref{lax2}). This allows one to identify these symplectic leaves with $T^\ast J$ and to treat  $p_1,\dots,p_g$ and $q_1,\dots,q_g$ as canonical coordinates on the symplectic leaves themselves.

In our case instead, it is not hard to realize that  the suitably de-singularised spectral curve $\Gamma$ is an hyperelliptic curve of genus $g=2 n$, so that ${\rm dim}(T^\ast J)=4n$, which is twice the dimension of our symplectic leaves. In fact the characteristic equation has the following form
$$
\mu^2=4^{2n} \lambda^{4n} + Pol_{2n-1}(\lambda^2) + \frac{\alpha_n^2}{\lambda^2},
$$
where $Pol_{2n-1}$ is a polynomial of degree $2n-1$. By doing if necessary a small monodromy preserving deformation, we can assume this polynomial to be irreducible. Setting $\tilde\mu=\lambda\mu$, we get
$$
\tilde\mu^2=Pol_{4n+2}(\lambda),
$$
where $Pol_{4n+2}$ is an irreducible polynomial in $\lambda$ of degree $4n+2$. We see that the genus is $2n$.

Another problem is that the coordinates $q_1,\dots,q_g$ are defined by taking the roots of the polynomial (\ref{eq:smallq}), so they may not satisfy the Painlev\'e property of the isomonodromic deformations equations (see \cite{malg,miwa}).  Therefore we propose a new set of canonical coordinates:

\begin{theorem}\label{mainth1}
Consider the following
$$
P_k = \Pi_{2k}=\frac{a^{(n)}_{2(n-k)+1}+b^{(n)}_{2(n-k)+1}}{a^{(n)}_{2n+1}},
\qquad  Q_k=\sum_{j=1}^n
\frac{1}{2 j} b^{(n)}_{2j}    \frac{\partial S_{2j}}{\partial\Pi_{2k}}, \qquad k=1,\dots,n,
$$
where $S_{k}=\sum_{j=1}^{2n} q_j^{k}$ for $k=1,\dots,2n$ and $\Pi_1,\dots,\Pi_{2n}$ are
the symmetric functions of $q_1,\dots, q_{2n}$:
$$
\Pi_1=q_1+q_2+\dots+q_{2n},\quad \Pi_2=\sum_{1\leq j< k\leq 2n}
q_j q_k,\quad \dots,\quad  \Pi_{2n} = q_1 q_2\dots q_{2n}.
$$
Then\begin{enumerate}
\item $P_1,\dots,P_n,Q_1,\dots,Q_n$ are coordinates in the symplectic leaves.
\item $P_1,\dots,P_n,Q_1,\dots,Q_n$ are canonical, namely
$$
\{P_i,P_j\}=\{Q_i,Q_j\}=0,\qquad
\{P_i,Q_j\}=\delta_{ij}.
$$
\end{enumerate}
\end{theorem}

\begin{remark}
Observe that the first statement of our theorem could have been guessed by noticing that, as in the case of the Kowalevski top,  $\Gamma$ admits one extra symmetry $\lambda\to-\lambda$ apart from the hyperelliptic involution $\mu\to-\mu$. As a consequence $\Gamma$ is a two-sheeted covering of  a genus $n$ hyperelliptic curve $C$ obtained by setting $z=\lambda^2$:
$$
C=\left\{\mu^2=4^{2n} z^{2n} +  Pol_{2n-1}(z) + \frac{\alpha_n^2}{z}
\right\}.
$$
Differently from the case of the Kowalevski top, this cover is branched at $0$ and at $\infty$.  Having fixed the normalization $c_1=-c_2=1$, we see that the Baker-Akhiezer function $\psi$ has a simple pole at $\lambda=\infty$ and $2n$ simple poles at $\lambda=q_1,\dots,q_{2n}$. This pole divisor is clearly invariant w.r.t. the involution $\lambda\to-\lambda$. In fact the poles $\lambda=q_1,\dots,q_{2n}$ come in pairs $q_j, q_{n+j}$ and each pair projects to one pole on $C$.
Due to the construction by A. Weil (see for example \cite{mum}),
the symmetric functions of the pole divisor on $C$ appear naturally as coordinates when endowing $Jac(C)$ with the structure of algebraic variety.\footnote{We are grateful to M. Talon for pointing out to us the similarity with the Kowalevski case and with the construction by A. Weil.}
\end{remark}

\begin{proof}
The first statement of the theorem follows as a straightforward corollary of Theorem \ref{thm:ABB} proved in Section \ref{se:expl} below.

Let us prove the second statement of our theorem, i.e. that our coordinates $P_1,\dots,P_n,Q_1,\dots,Q_n$ are canonical.

To compute the Poisson brackets between our coordinates, we observe that $\Pi_{2k+1}=S_{2k+1}=0$ in our case and
$$
\Pi_{2k}=\frac{a^{(n)}_{2(n-k)+1}+b^{(n)}_{2(n-k)+1}}{a^{(n)}_{2n+1}},\qquad
k=1,\dots,n.
$$
First let us compute the bracket $\{P_k,P_l\}$
\begin{eqnarray*}
\{P_k,P_l\}&=&\left\{\frac{a^{(n)}_{2(n-k)+1}+b^{(n)}_{2(n-k)+1}}{a^{(n)}_{2n+1}},
\frac{a^{(n)}_{2(n-l)+1}+b^{(n)}_{2(n-l)+1}}{a^{(n)}_{2n+1}}\right\}=\\
&=&\left({1\over
a^{(n)}_{2n+1}}\right)^2\left(\left\{b^{(n)}_{2(n-l)+1},a^{(n)}_{2(n-k)+1}\right\}-
\left\{b^{(n)}_{2(n-k)+1},a^{(n)}_{2(n-l)+1}\right\}\right)=\\
&=&\left({1\over
a^{(n)}_{2n+1}}\right)^2\left(b^{(n)}_{2(2n-l-k+1)}-b^{(n)}_{2(2n-l-k+1)}\right)=0.
\end{eqnarray*}
Therefore we have
\begin{eqnarray}\label{eq:pp}
\{P_k,P_l\}=\{\Pi_{2k},\Pi_{2l}\}=0, \qquad k,l=1,\ldots,n.
\end{eqnarray}
To compute the brackets that involve the $Q_k$, we make use of the
following formula:
\begin{eqnarray}\label{eq:gen}
\ln\left(\sum_{j=0}^{\infty}\Pi_j\gamma^j\right)=
\sum_{k=1}^{\infty}(-1)^{k-1}{S_{k}\over k}\gamma^k,
\end{eqnarray}
where $\gamma$ is an auxiliary variable.
Since in our case, the roots of the polynomial
$$
 \sum_{k=0}^{n-1}(b^{(n)}_{2k+1}+a^{(n)}_{2k+1}) \lambda^{2k}+
a^{(n)}_{2n+1}\lambda^{2n}=0
$$
are given by $q_1,\ldots,q_n,-q_1,\ldots,-q_n$, we have
$$
\Pi_{2k+1}=S_{2k+1}=0.
$$
Therefore, by differentiating (\ref{eq:gen}) with respect to
$\Pi_{2k}$, we can express $\frac{\partial
S_{2j}}{\partial\Pi_{2k}}$ as follows
\begin{eqnarray}\label{eq:diff}
\frac{\partial
S_{2j}}{\partial\Pi_{2k}}=-(2j)\left[\left(\sum_{i=0}^{n}\Pi_{2i}\gamma^{2i}\right)^{-1}\right]_{2j-2k},
\end{eqnarray}
where $[X(\gamma)]_{2j-2k}$ is the coefficient of $\gamma^{2j-2k}$ of
$X(\gamma)$ considered as a power series in $\gamma$ near $0$.
Note that although the sum in the left hand side of
(\ref{eq:gen}) goes from 1 to $\infty$, only terms where $j\leq n$
enter in (\ref{eq:diff}) as $j-k\leq n$.

We will now compute the bracket $\{P_k,Q_l\}$
\begin{eqnarray}\label{eq:pq}
\{P_k,Q_l\}=\sum_{j=1}^n\{\Pi_{2k},b_{2j}^{(n)}\}{1\over
2j}\frac{\partial
S_{2j}}{\partial\Pi_{2l}}+\sum_{j=1}^nb_{2j}^{(n)}{1\over
2j}\left\{\Pi_{2k},\frac{\partial
S_{2j}}{\partial\Pi_{2l}}\right\}.
\end{eqnarray}
Since $\frac{\partial S_{2j}}{\partial\Pi_{2l}}$ is a polynomial
in $\Pi_{2m}$ with $m\leq j$, the second term in (\ref{eq:pq}) is
zero because of (\ref{eq:pp}). The first term in (\ref{eq:pq}) is
\begin{eqnarray*}
\{P_k,Q_l\}&=&\sum_{j=1}^n\{\Pi_{2k},b_{2j}^{(n)}\}{1\over
2j}\frac{\partial S_{2j}}{\partial\Pi_{2l}}=\\
&=&\sum_{j=1}^n\Pi_{2(k-j)}\left[\left(\sum_{j=0}^{n}\Pi_{2j}\gamma^{2j}\right)^{-1}\right]_{2j-2l}=\\
&=&\sum_{j=l}^{n}\Pi_{2(k-j)}\left[\left(\sum_{j=0}^{n}\Pi_{2j}\gamma^{2j}\right)^{-1}\right]_{2j-2l},
\end{eqnarray*}
where we replaced the sum from 1 to $n$ by a sum from $l$ to $n$ in
the last equation because the expression
$$
\left(\sum_{j=0}^{n}\Pi_{2j}\gamma^{2j}\right)^{-1}
$$
does not contain any negative power. Since the expression
$$
\sum_{j=l}^{n}\Pi_{2(k-j)}\left[\left(\sum_{j=0}^{n}\Pi_{2j}\gamma^{2j}\right)^{-1}\right]_{2j-2l}
$$
is just the coefficient of $\gamma^{2k-2l}$ in
$$
\left(\sum_{i=0}^{n}\Pi_{2i}\gamma^{2i}\right)\left(\sum_{j=0}^{n}\Pi_{2j}\gamma^{2j}\right)^{-1}=1,
$$
we see that
$$
\{P_k,Q_l\}=\delta_{kl}.
$$
To compute the brackets between $Q_k$ and $Q_l$, once again, note
that since the brackets
\begin{eqnarray*}
\{b_{2j}^{(n)},b_{2i}^{(n)}\}=\left\{\frac{\partial
S_{2j}}{\partial\Pi_{2k}},\frac{\partial
S_{2i}}{\partial\Pi_{2l}}\right\}=0,
\end{eqnarray*}
the only contributions to the bracket $\{Q_k,Q_l\}$ come from the
cross terms
\begin{eqnarray}\label{eq:qq}
\{Q_k,Q_l\}&=&\sum_{j,i=1}^n\left\{\frac{1}{2 j} b^{(n)}_{2j}
\frac{\partial S_{2j}}{\partial\Pi_{2k}},\frac{1}{2 i}
b^{(n)}_{2i} \frac{\partial S_{2i}}{\partial\Pi_{2l}}\right\}\nn=\\
&=&\sum_{j,i=1}^n\frac{1}{4 ij} \left(b^{(n)}_{2j}
\left\{\frac{\partial
S_{2j}}{\partial\Pi_{2k}},b^{(n)}_{2i}\right\} \frac{\partial
S_{2i}}{\partial\Pi_{2l}}- b^{(n)}_{2i} \left\{\frac{\partial
S_{2i}}{\partial\Pi_{2l}},b^{(n)}_{2j}\right\} \frac{\partial
S_{2j}}{\partial\Pi_{2k}}\right)=\\
&=&\sum_{j,i=1}^n\frac{1}{4 ij} \left(b^{(n)}_{2j}
\left\{\frac{\partial
S_{2j}}{\partial\Pi_{2k}},b^{(n)}_{2i}\right\} \frac{\partial
S_{2i}}{\partial\Pi_{2l}}- b^{(n)}_{2j} \left\{\frac{\partial
S_{2j}}{\partial\Pi_{2l}},b^{(n)}_{2i}\right\} \frac{\partial
S_{2i}}{\partial\Pi_{2k}}\right).\nn
\end{eqnarray}
The bracket between $\frac{\partial S_{2j}}{\partial\Pi_{2k}}$ and
$b_{2i}^{(n)}$ can be computed as
\begin{eqnarray*}
\left\{\frac{\partial
S_{2j}}{\partial\Pi_{2k}},b_{2i}^{(n)}\right\}&=
&2j\left[\left(\sum_{m=0}^{n}\Pi_{2m}\gamma^{2m}\right)^{-2}
\sum_{s=0}^{n}\left\{\Pi_{2s},b^{(n)}_{2i}\right\}\gamma^{2s}\right]_{2j-2k}=\\
&=&-2j\left[\left(\sum_{m=0}^{n}\Pi_{2m}\gamma^{2m}\right)^{-2}
\sum_{s=0}^{n}\Pi_{2(s-i)}\gamma^{2s}\right]_{2j-2k}=\\
&=&-2j\left[\left(\sum_{m=0}^{n}\Pi_{2m}\gamma^{2m}\right)^{-2}
\sum_{s=0}^{n-i}\Pi_{2s}\gamma^{2s+2i}\right]_{2j-2k}=\\
&=&-2j\left[\left(\sum_{m=0}^{n}\Pi_{2m}\gamma^{2m}\right)^{-2}
\sum_{s=0}^{n-i}\Pi_{2s}\gamma^{2s}\right]_{2j-2k-2i}=\\
&=&-2j\left[\left(\sum_{m=0}^{n}\Pi_{2m}\gamma^{2m}\right)^{-2}
\sum_{s=0}^{n}\Pi_{2s}\gamma^{2s}\right]_{2j-2k-2i},
\end{eqnarray*}
where in the last line, we replaced the upper limit of the second
sum by $n$. This is because $2j-2k-2i\leq 2(n-i)$, so if $p>n-i$,
the coefficient of $\gamma^{2p}$ in $\sum_{s=0}^{n}\Pi_{2s}\gamma^{2s}$ will
not enter in the final expression.

Therefore we have
\begin{eqnarray*}
\sum_{i=1}^n{{1}\over {4ij}}\left\{\frac{\partial
S_{2j}}{\partial\Pi_{2k}},b^{(n)}_{2i}\right\} \frac{\partial
S_{2i}}{\partial\Pi_{2l}}=\left[\left(\sum_{m=0}^{n}\Pi_{2m}\gamma^{2m}\right)^{-3}
\sum_{s=0}^{n}\Pi_{2s}\gamma^{2s}\right]_{2j-2k-2l}.
\end{eqnarray*}
Similarly, the second term in (\ref{eq:qq}) is given by
\begin{eqnarray*}
\sum_{i=1}^n{1\over
{4ij}}\left\{\frac{\partial
S_{2j}}{\partial\Pi_{2l}},b^{(n)}_{2i}\right\} \frac{\partial
S_{2i}}{\partial\Pi_{2k}}=\left[\left(\sum_{m=0}^{n}\Pi_{2m}\gamma^{2m}\right)^{-3}
\sum_{s=0}^{n}\Pi_{2s}\gamma^{2s}\right]_{2j-2k-2l},
\end{eqnarray*}
therefore the first and second terms in (\ref{eq:qq}) equal each
other and we have $\{Q_k,Q_l\}=0$.

In summary, we have
\begin{eqnarray*}
\{P_k,Q_l\}=\delta_{kl},\quad \{P_k,P_l\}=0,\quad
\{Q_k,Q_l\}=0,
\end{eqnarray*}
as we wanted to prove.\end{proof}

\begin{example}
Case $n=1$. In this case
$Q_1=\frac{1}{2} b^{(1)}_2 \frac{\partial
S_{2}}{\partial\Pi_{2}}=-b^{(1)}_2$ and
$P_1=\Pi_2=\frac{a^{(1)}_1+b^{(1)}_1}{a^{(1)}_3}$, that is
$$
Q=4 w,\qquad P=\frac{1}{2}\left(w_z-w^2-\frac{z}{2}\right).
$$
These coincide with Okamoto's canonical coordinates (up to a constant factor) \cite{okam}.
\end{example}

\begin{example}
Case $n=2$. In this case  $P_1=\Pi_2=\frac{a^{(2)}_3+b^{(2)}_3}{a^{(2)}_5}$,
$P_2=\Pi_4=\frac{a^{(2)}_1+b^{(2)}_1}{a^{(2)}_5}$, $Q_1= \frac{1}{2} b^{(2)}_{2}
\frac{\partial S_{2}}{\partial\Pi_{2}}+
 \frac{1}{4} b^{(2)}_{4}    \frac{\partial S_{4}}{\partial\Pi_{2}}=
 -b^{(2)}_{2}+b^{(2)}_{4}  \Pi_2$,  and
$Q_2=  \frac{1}{4} b^{(2)}_{4}    \frac{\partial
S_{4}}{\partial\Pi_{4}}=- b^{(2)}_{4} $ so, finally
\begin{eqnarray}
&&
P_1=-\frac{1}{2}(w^2-w_z-\frac{ {t}_1}{2}),\nn\\
&&
P_2=\frac{1}{16}(-z+6 w^4-12 w^2
w_z +2 w_z^2-4 w w_{zz}+2 w_{zzz}+2 {t}_1(w_z-w^2)),\nn\\
&&
Q_1=-8 w w_z +4 w_{zz},\nn\\
&&Q_2=16 w.
\nn\end{eqnarray}

\end{example}

\section{The coefficients of ${\mathcal A}^{(n)}$ as polynomials in the canonical
coordinates}\label{se:expl}

This Section is completely devoted to the proof of the theorem below
which expresses the matrix $\mathcal A^{(n)}$ as a polynomial in
$P_1,\dots,P_n$, $Q_1,\dots,Q_n$ and $z,{t}_1,\dots,
t_{n-1}$.

\begin{theorem}\label{thm:ABB}
Let $\mathcal{A}$, $\mathcal{B}_{odd}$ and $\mathcal{B}_{even}$ be the following polynomials in $\lambda$:
\begin{eqnarray}\label{eq:matcoef}
\mathcal{A}=\sum_{i=0}^{n}a_{2i+1}^{(n)}\lambda^{2i+1}, \quad\mathcal{B}_{odd}=\sum_{i=0}^{n}b_{2i+1}^{(n)}\lambda^{2i+1},
\quad\mathcal{B}_{even}=\sum_{i=1}^{n}b_{2i}^{(n)}\lambda^{2i},
\end{eqnarray}
and $\mathcal{P}$, $\mathcal{Q}$ and $\mathcal{T}$  the following truncated series:
\begin{eqnarray}\label{eq:pqt}
{\mathcal Q}=\sum_{i=1}^{n}Q_i\lambda^{2i},\quad\mathcal
{P}=\sum_{i=1}^nP_i\lambda^{-2i},\quad\mathcal{T}&=&\sum_{i=1}^{n-1}{t}_i(2\lambda)^{2i-2n}-z(2\lambda)^{-2n}.
\end{eqnarray}
Then the following relations hold true:
\begin{subequations}\label{eq:ABB}
\begin{eqnarray}
\mathcal{A}&=&\left({1\over 4}(2\lambda)^{2n+1}\left(1+\mathcal{P}-
{{(1+\mathcal{T})^2}\over{1+\mathcal{P}}}\right)-(2\lambda)^{-2n-1}
\mathcal{Q}^2(1+\mathcal{P})\right)_+,\label{eq:Afinal}\\
& &\nn\\
\mathcal{B}_{odd}&=&\left({1\over 4}(2\lambda)^{2n+1}\left(1+\mathcal{P}-{{(1+\mathcal{T})^2}\over{1+\mathcal{P}}}\right)+(2\lambda)^{-2n-1}
\mathcal{Q}^2(1+\mathcal{P})\right)_+,\label{eq:Boddfinal}\\
& &\nn\\
\mathcal{B}_{even}&=&-\lambda^{2}\left(\lambda^{-2}\mathcal{Q}(1+\mathcal{P})\right)_+,
\label{eq:Beven}
\end{eqnarray}
\end{subequations}
where in the above, the inverse $(1+\mathcal{P})^{-1}$ is to be interpreted as:
\begin{eqnarray}\label{eq:inverse}
(1+\mathcal{P})^{-1}=\sum_{i=0}^{\infty}(-{\mathcal P})^i
\end{eqnarray}
and $X_+$ indicates the polynomial part of the Laurent series $X$.

In particular, by comparing the coefficients of $\lambda$ in (\ref{eq:ABB}), we can express the matrix entries $a_{2i+1}^{(n)}$, $b_{2i+1}^{(n)}$ and $b_{2i}^{(n)}$ as polynomials in the canonical coordinates and the times.
\end{theorem}

\begin{proof}
Throughout the proof of this theorem we shall use the following
facts: let  $X$, $X^{\prime}$ and $Y$, $Y^{\prime}$ be Laurent
series in $\lambda$ with no positive part, such that
\begin{eqnarray}\nn
(\lambda^{2n-1}X)_+&=&(\lambda^{2n-1}X^{\prime})_+\nonumber\\
(\lambda^{2n-1}Y)_+&=&(\lambda^{2n-1}Y^{\prime})_+\nn
\end{eqnarray}
then \begin{eqnarray}\label{eq:xy2}
(\lambda^{2n-1}XY)_+&=&(\lambda^{2n-1}X^{\prime}Y^{\prime})_+,
\quad\hbox{and}\quad
(\lambda^{2n-1}X^{-1})_+=(\lambda^{2n-1}(X^{\prime})^{-1})_+.
\end{eqnarray}

Let us prove (\ref{eq:Beven}) first. Using the definition
of $Q_k$ given in Theorem \ref{mainth1},  we have
\begin{eqnarray*}
b^{(n)}_{2k}&=&\frac{-1}{2 k} b^{(n)}_{2k}    \frac{\partial
S_{2k}}{\partial\Pi_{2k}}=\\&=&-Q_k+{1\over
{2k+2}}b^{(n)}_{2k+2} \frac{\partial
S_{2k+2}}{\partial\Pi_{2k}}+{1\over {2k+4}}b^{(n)}_{2k+4}
\frac{\partial S_{2k+4}}{\partial\Pi_{2k}}+\cdots
\end{eqnarray*}
now apply the above repeatedly
to $b^{(n)}_{2k+2l}$ to obtain the following
\begin{eqnarray}\label{eq:b2n}
 b^{(n)}_{2k}
&=&-Q_k-Q_{k+1}\left({1\over {2k+2}} \frac{\partial
S_{2k+2}}{\partial\Pi_{2k}}\right)-\nn\\
&-&Q_{k+2}\left({1\over
{2k+4}} \frac{\partial S_{2k+4}}{\partial\Pi_{2k}}+{1\over
{2k+4}}{1\over {2k+2}}\frac{\partial
S_{2k+4}}{\partial\Pi_{2k+2}}\frac{\partial
S_{2k+2}}{\partial\Pi_{2k}}\right)-\\
&-&\cdots-Q_{k+l}\left({1\over
{2k+2l}}\frac{\partial S_{2k+2l}}{\partial\Pi_{2k}}+
\sum_{m=1}^{\infty}U_{2k+2l}^m\right)+\cdots\nn
\end{eqnarray}
where
\begin{eqnarray}\label{eq:R2k}
U_{2k+2l}^m&=&\sum_{j_1=m}^{l-1}{1\over {2k+2l}}\frac{\partial
S_{2k+2l}}{\partial\Pi_{2k+2j_1}}\sum_{j_2=m-1}^{j_1-1}{1\over
{2k+2j_1}}\frac{\partial
S_{2k+2j_1}}{\partial\Pi_{2k+2j_2}}\\&\cdots&\sum_{j_m=1}^{j_{m-1}-1}{1\over {2k+2j_{m-1}}}\frac{\partial
S_{2k+2j_{m-1}}}{\partial\Pi_{2k+2j_m}}{1\over
{2k+2j_m}}\frac{\partial S_{2k+2j_{m}}}{\partial\Pi_{2k}}.\nonumber
\end{eqnarray}
Note that, by (\ref{eq:diff}), we have
\begin{eqnarray}\label{eq:diff2}
\frac{\partial
S_{2p}}{\partial\Pi_{2q}}=
-(2p)\left[\left(\sum_{j=0}^{n}\Pi_{2j}(\gamma)^{2j}\right)^{-1}\right]_{2p-2q}.
\end{eqnarray}
This allows us to interpret the $U_{2k+2l}^m$ as coefficients of an
infinite series, which is a product of the series in the right hand
side of (\ref{eq:diff2}). However, we need to be careful since the
sums in (\ref{eq:R2k}) begin with terms that correspond to the
coefficient of $\gamma^2$ in (\ref{eq:diff2}). This means that
\begin{eqnarray*}
U_{2k+2l}^m=\textrm{ coefficient of $\gamma^{2l}$ in
$\left(1+\left(-\sum_{j=0}^{n}\Pi_{2j}\gamma^{2j}\right)^{-1}\right)^{m+1}$.}
\end{eqnarray*}
We can now identify $Q_k$ as coefficients of $\lambda$ in the polynomial $\mathcal{Q}$, $b_{2k}^{(n)}$ as coefficients of $\lambda$ in the polynomial $\mathcal{B}_{even}$. If we now substitute $\gamma=\lambda^{-1}$ and interpret $U_{2k+2l}^m$ also as coefficients of $\lambda$ in the product series, we can rewrite (\ref{eq:R2k}) as follows
\begin{eqnarray*}
\mathcal{B}_{even}&=&
-\lambda^2\left(\lambda^{-2}\mathcal{Q}\sum_{m=0}^{\infty}
\left(1+\left(-\sum_{j=0}^{n}\Pi_{2j}\lambda^{-2j}\right)^{-1}\right)^m\right)_+=\nonumber\\
&=&-\lambda^2\left(\lambda^{-2}\mathcal{Q}
\left(\left(\sum_{j=0}^{n}\Pi_{2j}\lambda^{-2j}\right)^{-1}\right)^{-1}\right)_+=\nonumber\\
&=&-\lambda^2\left(\lambda^{-2}\mathcal{Q}
\left(\sum_{j=0}^{n}\Pi_{2j}\lambda^{-2j}\right)\right)_+=\nonumber\\
&=&-\lambda^2\left(\lambda^{-2}\mathcal{Q}(1+\mathcal{P})\right)_+,
\end{eqnarray*}
which proves (\ref{eq:Beven}).

Let us now prove (\ref{eq:Boddfinal}). Notice that in the expressions (\ref{isomonodromy.3}) for $b_{2k+1}^{(n)}$ the Lenard operators appear together with their first and second derivatives. The canonical coordinates instead involve only the Lenard operators and their first derivatives. In fact,
the $P_1,\dots,P_n$ are expressed in terms of sums $a^{(n)}_{2j+1} + b^{(n)}_{2j+1}$ which depend only on the Lenard operators without derivatives, and the $Q_1,\dots,Q_n$ depend on the same sums and on the even coefficients $b_{2k}^{(n)}$ which are only expressed in terms of the Lenard operators and their first derivatives.

Therefore, to express the odd coefficients $b_{2k+1}^{(n)}$ in terms of the canonical coordinates, we will need to express the second derivatives of the Lenard operators in terms of the Lenard operators themselves and their first derivatives.  In terms of the generating function, this is given by
\begin{proposition}
Let ${\mathcal L}$ be the generating function of the Lenard recursion operator:
\begin{eqnarray*}
{\mathcal L}&:=&\sum_{i=1}^{\infty}{\mathcal L}_{i}\xi^{i},
\end{eqnarray*}
where $\xi$ is an auxiliary variable. Then the following relation holds true:
\begin{eqnarray}\label{eq:lzzcpt}
\p_z^2{\mathcal L}&=
&{1\over 2}(\xi^{-1}{\mathcal L}-{\mathcal L}_1)-2{\mathcal L}{\mathcal L}_1\\
&+&(1+2{\mathcal L})^{-1}\left({1\over
2}(\xi^{-1}{\mathcal L}-{\mathcal L}_1)-{\mathcal L}{\mathcal L}_1
+(\p_z{\mathcal L})^2\right).\nonumber
\end{eqnarray}
In particular, the above equation expresses the second derivatives of the Lenard operators as a polynomial in the Lenard operators and their first derivatives.
\end{proposition}

\begin{proof} We prove (\ref{eq:lzzcpt}) at each order in $\xi$. At order $\xi$,  the equation
(\ref{eq:lzzcpt}) is trivially satisfied. At order $\xi^n$, $n>0$, by multiplying (\ref{eq:lzzcpt}) by $(1+2{\mathcal L})$,  we get:
\begin{eqnarray}\label{eq:lzz}
{\mathcal L}_{n+1} &=&
\p_z^2{\mathcal L}_{n}+3{\mathcal
L}_{n}{\mathcal L}_{1}+\sum_{j=1}^{n-1}\Bigg(4{\mathcal L}_{1}
{\mathcal L}_{j}{\mathcal L}_{n-j}- \p_z{\mathcal
L}_{j}\p_z{\mathcal L}_{n-j} -{\mathcal L}_{j+1}{\mathcal
L}_{n-j}-2{\mathcal L}_{n-j}\p_z^2{\mathcal L}_{j}\Bigg) .
\end{eqnarray}
To prove this, we integrate relation (\ref{lenard}) by parts iteratively.
At the first step, we get:
\begin{eqnarray}\label{eq:Ln}
{\mathcal L}_{n+1}=
\left(\p_z^2+2{\mathcal L}_{1}\right){{\mathcal L}}_n
+2\int{\mathcal L}_{1}\p_z{{\mathcal L}}_n\d z,
\end{eqnarray}
where we have replaced $w_z-w^2$ by ${\mathcal L}_1$. It
is therefore sufficient to compute
$\int{\mathcal L}_{1}\p_z{{\mathcal L}}_n\d z$.
To achieve this, we will now compute a more general term
$\int{\mathcal L}_{i}\p_z{{\mathcal L}}_k\d z$
for any $k$ and $i$. First let us replace the term
$\p_z{{\mathcal L}}_k$ by using (\ref{lenard}):
\begin{eqnarray*}
\int{\mathcal L}_{i}\p_z{{\mathcal L}}_k\d
z=\int{\mathcal L}_i
\left(\p_z^3+4{\mathcal L}_{1}\p_z+
2({\mathcal L}_{1})_z\right){\mathcal L}_{k-1}\d z.
\end{eqnarray*}
We now integrate the first and last terms by parts, where the
integration by parts in the last term is performed as follows, let
\begin{eqnarray*}
F(z)&=&2\int{\mathcal L}_{i}({\mathcal L}_{1})_z\d
z\\
&=&\p_z^2{\mathcal L}_{i}+4{\mathcal L}_{1}{\mathcal L}_{i}-{\mathcal L}_{i+1},
\end{eqnarray*}
where we used the Lenard recursion relation (\ref{lenard}) to
obtain the above equality. We then have:
\begin{eqnarray*}
2\int{\mathcal L}_{i}({\mathcal L}_{1})_z{\mathcal L}_{k-1}\d
z&=&F(z){\mathcal L}_{k-1}- \int
F(z)\p_z{\mathcal L}_{k-1}\d z.
\end{eqnarray*}
>From this we have
\begin{eqnarray*}
\int{\mathcal L}_{i}\p_z{{\mathcal L}}_k\d
z&=&{\mathcal L}_i\p_z^2{\mathcal L}_{k-1}-\p_z{\mathcal L}_i\p_z{\mathcal L}_{k-1}
+{\mathcal L}_{k-1}\p_z^2{\mathcal L}_i+
4{\mathcal L}_1{\mathcal L}_i{\mathcal L}_{k-1}\\
&-&{\mathcal L}_{i+1}{\mathcal L}_{k-1}+\int
{\mathcal L}_{i+1}\p_z{\mathcal L}_{k-1}\d z
\end{eqnarray*}
We can replace the term $\int
{\mathcal L}_{i+1}\p_z{\mathcal L}_{k-1}\d z$ on
the right hand side by similar formula and express $\int
{\mathcal L}_{i}\p_z{\mathcal L}_{k}\d z$ as a
polynomial in ${\mathcal L}_{j}$,
$\p_z{\mathcal L}_{j}$ and
$\p_z^2{\mathcal L}_{j}$ for $j<k$. In particular, we
have
\begin{eqnarray*}
2\int{\mathcal L}_{1}\p_z{{\mathcal L}}_n\d
z&=&{\mathcal L}_{1}{\mathcal L}_{n}+\sum_{j=1}^{n-1}\Bigg(4{\mathcal L}_{1}
{\mathcal L}_{j}{\mathcal L}_{n-j}-
\p_z{\mathcal L}_{j}\p_z{\mathcal L}_{n-j}-\\
&-&{\mathcal L}_{j+1}{\mathcal L}_{n-j}+2{\mathcal L}_{j}\p_z^2{\mathcal L}_{n-j}\Bigg).
\end{eqnarray*}
By substituting this into (\ref{eq:Ln}), we finally get (\ref{eq:lzz}).
 \end{proof}

We can now express the odd coefficients $b_{2k+1}^{(n)}$ in terms of the canonical coordinates.

If we make the substitution $\xi={1\over 4}\lambda^{-2}$ in the generating function ${\mathcal L}$, then by (\ref{isomonodromy.3}), we have
\begin{eqnarray}\label{eq:oddeven}
\mathcal{B}_{odd}&=&\left((2\lambda)^{2n-1}\left({1\over
2}\p_z^2{\mathcal L}+w\p_z{\mathcal L}+
w_z({\mathcal L}+{\mathcal L}_0)\right)(1+\mathcal{T})\right)_+,\\
\mathcal{B}_{even}&=&(2\lambda)^2\left((2\lambda)^{2n-2}\left(-\p_z{\mathcal L}-2w({\mathcal L}+{\mathcal L}_0)\right)(1+\mathcal{T})\right)_+.\\
\end{eqnarray}
Note that, since
\begin{eqnarray*}
\left((2\lambda)^{2n-1}\p_z{\mathcal L}(1+\mathcal{T})\right)_+
\end{eqnarray*}
is a series in odd powers of $\lambda$ only, while
\begin{eqnarray*}
\left((2\lambda)^{2n-2}\p_z{\mathcal L}(1+\mathcal{T})\right)_+
\end{eqnarray*}
is a series in even powers of $\lambda$ only, we have
\begin{eqnarray*}
2\lambda\left((2\lambda)^{2n-2}\p_z{\mathcal L}(1+\mathcal{T})\right)_+
=\left((2\lambda)^{2n-1}\p_z{\mathcal L}(1+\mathcal{T})\right)_+.
\end{eqnarray*}
>From this, and the second equation in (\ref{eq:oddeven}), we can express the second term in $\mathcal{B}_{odd}$ in terms of $\mathcal{B}_{even}$
\begin{eqnarray*}
\left((2\lambda)^{2n-1}\p_z{\mathcal L}(1+\mathcal{T})\right)_+=
(2\lambda)^{-1}\mathcal{B}_{even}+
\left((2\lambda)^{2n-1}(2\mathcal{L}+1)w(1+\mathcal{T})\right)_+.
\end{eqnarray*}
By substituting this into the right hand side of $\mathcal{B}_{odd}$, we have
\begin{eqnarray*}
\mathcal{B}_{odd}&=&\left((2\lambda)^{2n-1}
\left({1\over 2}\p_z^2{\mathcal L}-w^2({\mathcal L}+{\mathcal L}_0)
+{\mathcal L}_1({\mathcal L}+{\mathcal L}_0)\right)(1+\mathcal{T})-(2\lambda)^{-1}\mathcal{B}_{even}\right)_+,\\
\mathcal{B}_{even}&=&(2\lambda)^2\left((2\lambda)^{2n-2}\left(-\p_z{\mathcal L}-2w{\mathcal L}-w+\right)(1+\mathcal{T})\right)_+.
\end{eqnarray*}
By substituting this into (\ref{eq:lzzcpt}), we obtain
\begin{eqnarray}\label{eq:Bodd}
\mathcal{B}_{odd}&=
&{1\over 2}\Bigg((2\lambda)^{2n-1}\Bigg[-2w^2{\mathcal L}-w^2
+{1\over 2}((2\lambda)^{2}{\mathcal L}+{\mathcal L}_1)+\nonumber\\
&+&(1+2{\mathcal L})^{-1}\Bigg({1\over
2}\left((2\lambda)^{2}{\mathcal L}-{\mathcal L}_1\right)-{\mathcal L}{\mathcal L}_1+\\
&+&(\p_z{\mathcal L})^2\Bigg)\Bigg](1+\mathcal{T})-2w(2\lambda)^{-1}\mathcal{B}_{even}\Bigg)_+.\nonumber
\end{eqnarray}
To complete the calculation, we need to express
${\mathcal L}$ in terms of $\mathcal{P}$. Since we have
\begin{eqnarray*}
P_{k}=\frac{a^{(n)}_{2(n-k)+1}+b^{(n)}_{2(n-k)+1}}{a^{(n)}_{2n+1}}
=\frac{2}{4^k}\sum_{l=n-k}^n{t}_l{\mathcal L}_{k+(l-n)} ,\qquad k=1,\dots,n,
\end{eqnarray*}
where ${t}_n=1$, we see that
\begin{eqnarray*}
{\mathcal L}_k&=&2^{2k-1}P_k-{t}_{n-1}{\mathcal L}_{k-1}+\cdots-{t}_{n-i}{\mathcal L}_{k-i}+\cdots,\\
{\mathcal L}_k&=&2^{2k-1}P_k+2^{2k-3}P_{k-1}(-{t}_{n-1})+\\
&+&2^{2(k-i)-1}P_{k-i}\left(-{t}_{n-i}+\cdots+\sum_{j_1=1}^{n-i-1}(-{t}_{n-i-j_1})
\sum_{j_2=1}(-{t}_{j_1-j_2})\cdots\sum_{j_l=1}^{j_{l-1}-1}(-{t}_{j_l})+\ldots\right)+\\
&+&{\mathcal L}_0\left({t}_{n-k}+\cdots+\sum_{j_1=1}^{n-k-1}(-{t}_{n-k-j_1})
\sum_{j_2=1}(-{t}_{j_1-j_2})\cdots\sum_{j_l=1}^{j_{l-1}-1}(-{t}_{j_l})+\ldots\right).
\end{eqnarray*}
Finally, by using similar argument as before, we see that
\begin{eqnarray}\label{eq:LP}
\left((2\lambda)^{2n-1}({\mathcal L}+{\mathcal L}_0)\right)_+=\left((2\lambda)^{2n-1}\left({\mathcal L}_0+{1\over
2}\mathcal{P}\right)(1+\mathcal{T})^{-1}\right)_+,
\end{eqnarray}
where $(1+\mathcal{T})^{-1}$ in the above is to be interpreted as in (\ref{eq:inverse}).
Thanks to (\ref{eq:xy2}), this implies the following
\begin{eqnarray}\label{eq:LP2}
\left((2\lambda)^{2n-1}({\mathcal L}+{\mathcal L}_0)^{-1}\right)_+=\left((2\lambda)^{2n-1}\left({\mathcal L}_0+{1\over
2}\mathcal{P}\right)^{-1}(1+\mathcal{T})\right)_+
\end{eqnarray}
By substituting (\ref{eq:LP}) and (\ref{eq:LP2}) into (\ref{eq:Bodd}), we get(\ref{eq:Boddfinal}).

To finish the proof of the theorem, note that by the definition of the coordinates $P_1,\dots,P_n$ in Theorem \ref{mainth1}, we have
\begin{eqnarray}\label{eq:ABP}
&&
\lambda(2\lambda)^{2n}(1+\mathcal{P})=(\mathcal{A}+\mathcal{B}_{odd}) \quad \Rightarrow\nn\\
&& \nn\\
&&
\mathcal{A}=(-\mathcal{B}_{odd}+\lambda(2\lambda)^{2n}(1+\mathcal{P}))
\end{eqnarray}
The proves the theorem. \end{proof}

\begin{example}
Let us illustrate how formulae (\ref{eq:Afinal}), (\ref{eq:Boddfinal}) and (\ref{eq:Beven}) work in a concrete but non trivial example. In the case $n=3$ the definitions  (\ref{eq:matcoef}) and (\ref{eq:pqt}) read:
\begin{eqnarray}\label{ex:n3}\nn
&&
{\mathcal A} = a^{(3)}_1 \lambda +  a^{(3)}_3 \lambda^3
+ a^{(3)}_5 \lambda^5+a^{(3)}_7\lambda^7,\nn\\
&&
{\mathcal B_{odd}} = b^{(3)}_1 \lambda +  b^{(3)}_3 \lambda^3+ b^{(3)}_5 \lambda^5,\nn\\
&&
{\mathcal B_{even}} = b^{(3)}_2 \lambda^2 +  b^{(3)}_4 \lambda^4+ b^{(3)}_6 \lambda^6,\nn\\
&&
{\mathcal Q}=Q_1 \lambda^2 + Q_2 \lambda^4 + Q_3 \lambda^6,\\
&&
{\mathcal P}= \frac{P_1}{\lambda^2} +\frac{P_2}{\lambda^4} +\frac{P_3}{\lambda^6},\nn\\
&&
{\mathcal T}=\frac{{t}_2}{(2\lambda)^2}+\frac{{t}_1}{(2\lambda)^4}-\frac{z}{(2 \lambda)^6}.
\nn
\end{eqnarray}
Theorem \ref{thm:ABB} allows to express the coefficients $a^{(3)}_1,a^{(3)}_3,a^{(3)}_5,a^{(3)}_7$
and the coefficients $b^{(3)}_1,\dots,b^{(3)}_6$ in terms of $P_1,P_2,P_3,Q_1,Q_2,Q_3$ and
$z,{t}_1,{t}_2$. Let us  use (\ref{eq:Beven}) first:
$$
{\mathcal Q}(1+\mathcal P) =\frac{P_3 Q_1}{\lambda^4}+\frac{P_2 Q_1+P_3 Q_2}{\lambda^4}+
P_1 Q_1+P_2 Q_2+P_3 Q_3 +\lambda^2(Q_1+P_1 Q_2+P_2 Q_3)+\lambda^4 (Q_2+P_1 Q_3)+
\lambda^6 Q_3,
$$
and after dividing by $\lambda^2$ and throwing away all negative powers, we get
$$
\left(\lambda^{-2}{\mathcal Q}(1+\mathcal P)\right)_+ =Q_1+P_1 Q_2+P_2 Q_3+
\lambda^2 (Q_2+P_1 Q_3)+ \lambda^4 Q_3.
$$
Now, we need to multiply by $\lambda^2$ again and to compare with $\mathcal B_{even}$. We get
$$
b^{(3)}_2= -Q_1-P_1 Q_2 -P_2 Q_3,\qquad
b^{(3)}_4 = -Q_2-P_1 Q_3,\qquad
b^{(3)}_6= -Q_3.
$$
Let us briefly illustrate how to obtain the other coefficients. The procedure is the same as above with the only complication of the term
$$
\left(
\frac{1}{4}(2\lambda)^7 \frac{1+\mathcal T}{1+\mathcal P}
\right)_+.
$$
Let us compute this term explicitly. Since we have a power $7$ in $\lambda$ in the numerator,
we need to only the terms up to order $\frac{1}{\lambda^7}$ in the expansion of
$({1+\mathcal P})^{-1}$ at $\infty$:
$$
({1+\mathcal P})^{-1}=1-\frac{P_1}{\lambda^2}+\frac{P_1^2-P_2}{\lambda^4}+
\frac{2 P_1P_2-P_3-P_1^3}{\lambda^6}+{\mathcal O}(\lambda^8).
$$
In this way we see that
\begin{eqnarray}\nn
&&
\left(\frac{1}{4}(2\lambda)^7 \frac{1+\mathcal T}{1+\mathcal P}\right)_+=
(64 P_1-16 {t}_2)\lambda^5 +
(64 P_2-32 P_1^2-4 {t}_1+16 {t}_2 P_1-2 {t}_2^2)\lambda^3+\nn\\
&&
\qquad
+\lambda(64 P_3+z+32 P_1^3-64 P_1 P_2+4 {t}_1 P_1-16 {t}_2 P_1^2+16 {t}_2 P_2-{t}_1{t}_2 + 2 {t}_2^2 P_1). \nn
\end{eqnarray}
Analogously, one can compute the other terms in (\ref{eq:Afinal}) and (\ref{eq:Boddfinal}) to obtain:
\begin{eqnarray}\nn
&&
b^{(3)}_1=z+32 P_1^3-64 P_1 P_2 + 64 P_3+\frac{Q_2^2}{128}+\frac{Q_1 Q_3}{64}+
\frac{P_1 Q_2 Q_3}{64} +\nn\\
&&
\quad\qquad+\frac{P_2 Q_3^2}{128} +4 {t}_1 P_1- 16 {t}_2 P_1^2+
16 {t}_2 P_2 -{t}_1 {t}_2 +2 {t}_2^2 P_1,\nn\\
&&
b^{(3)}_3= 64 P_2-32 P_1^2+\frac{Q_2 Q_3}{64}+
\frac{P_1Q_3^2}{128} - 4 {t}_1+16 {t}_2 P_1 -2 {t}_2^2,\nn\\
&&
b^{(3)}_5 =64 P_1 +\frac{Q_3^2}{128}-16 {t}_2,\nn\\
&&
a^{(3)}_1=z+32 P_1^3-64 P_1 P_2 +\frac{Q_2^2}{128}+\frac{Q_1 Q_3}{64}+
\frac{P_1 Q_2 Q_3}{64} +\nn\\
&&
\quad\qquad+\frac{P_2 Q_3^2}{128} +4 {t}_1 P_1- 16 {t}_2 P_1^2+
16 {t}_2 P_2 -{t}_1 {t}_2 +2 {t}_2^2 P_1,\nn\\
&&
a^{(3)}_3= -32 P_1^2+\frac{Q_2 Q_3}{64}+
\frac{P_1Q_3^2}{128} - 4 {t}_1+16 {t}_2 P_1 -2 {t}_2^2,\nn\\
&&
a^{(3)}_5 =+\frac{Q_3^2}{128}-16 {t}_2,
\qquad a^{(3)}_7 =64.
\nn
\end{eqnarray}
It is clear that we have only used linear algebra to obtain these coefficients, instead of using differentiation, integration and recursion.
\end{example}

\section{Hamiltonians}\label{se:ham}

In Section \ref{sec:coaj} we proved that the equation (\ref{isomonodromy.4}) is the same as (\ref{lax2}):
$$
\left(\partial_z-\partial_z^w\right) A={\rm ad}^\ast_B A,
$$
where $A=\left(\mathcal A^{(n)}\right)_+\in{\mathfrak g}_-^\ast$ is
the dynamical part of $\mathcal A^{(n)}$, and
$B=\left(\frac{\mathcal
A^{(n)}\lambda^{1-2n}}{4^n}\right)_-\in{\mathfrak g}_-$. This allows
us to interpret the evolution along $(\partial_z-\partial_z^w) $ as
a vector field on a coadjoint orbit of the twisted loop algebra
$\mathfrak g_-$. We are now going to show that this vector field is
Hamiltonian and that the isomonodromic deformation Hamiltonian for
the $n$-th equation in the PII hierarchy is given by
\begin{equation}\label{eq:ham}
H^{(n)}:= -\frac{1}{2\, 4^n} \tr\res\left(\lambda^{1-2n} \left(\A^{(n)}\right)^2\right).
\end{equation}
This fact is actually a consequence of a more general result:

\begin{proposition}\label{prop:ham}
The vector field
$$
{\mathcal X}_k(A):=-\left[\left(\mathcal A^{(n)}\lambda^{1-2k}\right)_-,A\right],
$$
is Hamiltonian with the Hamiltonian function
\begin{equation}\label{eq:ham1}
h^{(n)}_k:= \frac{1}{2} \tr\res\left(\lambda^{1-2k} \left(\A^{(n)}\right)^2\right).
\end{equation}
\end{proposition}

\begin{proof}
Let us denote
$$
\hat L_k = -\left({\mathcal A^{(n)}\lambda^{1-2k}}\right)_-\in{\mathfrak g}_-.
$$
We are interested in the vector field
$$
{\mathcal X}_k(A)=\left[\hat L_k,A\right].
$$
To show that it is Hamiltonian and to compute the Hamiltonian function $f$, we use the following definition:
$$
\omega({\mathcal X}_k,Y)(\Xi):=-\left<[Y,\Xi],\d f\right>, \quad Y\in{\frak g}_-,\,\Xi\in\mathfrak g_-^\ast,
$$
so that
$$
\omega({\mathcal X}_k,Y)(A)=-\left<\left[\hat L_k,Y\right],A\right>=\left<[Y,A],-\hat L_k\right>=<[Y,A],\d f>.
$$
This shows that if we can prove that there exist $f$ such that $\d f=-\hat L_k$, then $[\hat L_k,A]$ defines a Hamiltonian vector field of Hamiltonian $f$.

We are now going to show that the Hamiltonian (\ref{eq:ham1}) is
such that $\d h^{(n)}_k=-\hat L_k$. For every $X\in\mathfrak g_-$ and
$\Xi\in\mathfrak g_-^\ast$,  we can identify $[X,\Xi]={\rm
ad}^\ast_X \Xi$ with a vector tangent to the coadjoint orbit. Denote
this vector by $\delta_X \Xi$.  Let $\delta_{X} A\in T_A{\mathcal
O}_A$ then using the definition \ref{diff}, we get
\begin{eqnarray*}
h^{(n)}_k(A+\delta_{X} A)-h^{(n)}_k(A) +\mathcal O(\delta_{X} A)^2=
{1\over {2}}\left<\lambda^{1-2k}A, (2\delta_{X}\A^{(n)})\right>=\left<\delta_{X}A,\d
h^{(n)}_k\right>,
\end{eqnarray*}
which is the contraction between
$\delta_{X}A$ and $\d h^{(n)}_k$, as we wanted to prove.
\end{proof}

We now compute the Hamiltonian $\mathcal H^{(n)}$ in terms of the canonical coordinates.

\begin{theorem}
Define
\begin{equation}\label{hamPQ}
{\mathcal H}^{(n)}(P_1,\dots,P_n,Q_1,\dots,Q_n,z):= -\frac{1}{4^n} \left( \sum_{l=0}^{n-1} a^{(n)}_{2l+1}
a^{(n)}_{2(n-l)-1}-\sum_{l=0}^{n-1} b^{(n)}_{2l+1} b^{(n)}_{2(n-l)-1}+
\sum_{l=0}^{n} b^{(n)}_{2l} b^{(n)}_{2(n-l)}\right)+\frac{Q_n}{4^n},
\end{equation}
in which we are thinking of $a^{(n)}_{2l+1},b^{(n)}_{2l+1}, b^{(n)}_{2l}$ as the functions of $P_1,\dots,P_n,Q_1,\dots,Q_n,{t}_1,\dots,{t}_{n-1},z$ given by (\ref{eq:ABB}). Then the $n$-th member of the second \Pa\/  hierarchy is given
by the equations
\begin{equation}\label{eq:ham:PQ}
\frac{{\rm d}P_k}{{\rm d}z} = -
\frac{\partial \mathcal H^{(n)}}{\partial Q_k} ,\qquad
\frac{{\rm d}Q_k}{{\rm d}z} =
\frac{\partial \mathcal H^{(n)}}{\partial P_k}.
\end{equation}
\end{theorem}

\begin{proof}
Observe that equation (\ref{eq:ham}) gives
$$
H^{(n)}= -\frac{1}{4^n} \left( \sum_{l=0}^{n-1} a^{(n)}_{2l+1}
a^{(n)}_{2(n-l)-1}-\sum_{l=0}^{n-1} b^{(n)}_{2l+1} b^{(n)}_{2(n-l)-1}+
\sum_{l=0}^{n} b^{(n)}_{2l} b^{(n)}_{2(n-l)}
\right),
$$
where we are treating $a^{(n)}_j, b^{(n)}_j$ as coordinates on the coadjoint orbit.
When expressing this Hamiltonian in our canonical coordinates, we need to take into account a shift  $h$ due to the explicit $z$ dependence in the variable $P_n$. All other canonical coordinates depend on $a^{(n)}_j, b^{(n)}_j$ only.  To compute this shift we use the following well-known result:

\begin{lemma}\label{le:bo}
Let
\begin{equation}\label{hamy}
\frac{{\rm d}y_i}{{\rm d} z}=\{y_i,H(y,z)\},
\end{equation}
be a Hamiltonian system on a Poisson manifold with Poisson brackets $\{\cdot,\cdot\}$ and
$$
y=\phi(x,z),
$$
be a local diffeomorphism depending explicitly on $z$. Let the vector field $\partial_z                                                                                                        \phi$ be a Hamiltonian vector field with Hamiltonian $\delta H$. Then (\ref{hamy}) is a Hamiltonian system also in the $x$--coordinates
$$
\frac{{\rm d}x_i}{{\rm d} z}=\{x_i,\hat H(x,z)\},
$$
where
$$
\hat H(x,z)=H(\phi(x,z),z) -\delta H(\phi(x,z),z).
$$
\end{lemma}

Let us compute this shift in our case. The only coordinate depending explicitly on $z$ is
$P_n=\frac{-z}{4^n}+f(a^{(n)}_1,\dots,a^{(n)}_{2n+1}, b^{(n)}_0,b^{(n)}_1,\dots, b^{(n)}_{2n})$. So for $y=(P_1,\dots,P_n,Q_1,\dots,Q_n)$, we have $\delta H^{(n)}=\frac{Q_n}{4^n}$ which gives (\ref{hamPQ}).
\end{proof}

\begin{remark}
It is clear that the Hamiltonian equations (\ref{eq:ham:PQ}) satisfy the Painlev\'e property. In fact
(\ref{comp-ii}) satisfies the Painlev\'e property \cite{malg,miwa}, and since $\mathcal A^{(n)}$ is a polynomial in $P_1,\dots,P_n,Q_1,\dots,Q_n$, no movable critical points are introduced.
\end{remark}

\begin{example} In the case $n=1$, we get the $\Pt$ Hamiltonian \cite{okam,JM}
$$
\mathcal H^{(1)}= 4 P^2 +\frac{1}{4} Q+ \frac{1}{4} P Q^2+2 P z-\frac{1}{2} Q \alpha_1.
$$
The Hamilton's equations
$$
\frac{{\rm d}^2 Q}{{\rm d} z^2} = \frac{1}{8} Q^3+ Q z+4 \alpha_1,
$$
give the second \Pa\/ equation for $w=\frac{1}{4}Q$:
$$
\frac{{\rm d}^2 w}{{\rm d} z^2}=2 w^3+ w z +\alpha_1.
$$
\end{example}

\section{Higher order flows as time-dependent Hamiltonian systems}\label{se:time}

As illustrated in Section \ref{se:pii}, the time ${t}_k$ dependence is described by the rescaled mKdV equation (\ref{timeflows}), which is equivalent to the compatibility equation (\ref{eq-MB}). In the Appendix
\ref{app-a}, we proved that all equations (\ref{comp-ii}),  (\ref{eq-MB}) and ({\ref{laxMA}) are consistent, so that they indeed define isomonodromic deformations.
In this Section, we shall deduce the Hamiltonian functions $\mathcal H^{(n)}_k$ such that equation (\ref{timeflows}) is equivalent to
\begin{equation}\label{st-ham}
\frac{\partial Q_i}{\partial{t}_k}=\frac{\partial\mathcal  H^{(n)}_k}{\partial P_i},
\qquad
\frac{\partial P_i}{\partial{t}_k}=-\frac{\partial\mathcal  H^{(n)}_k}{\partial Q_i},
\quad i=1,\dots,n.
\end{equation}
Since we are dealing with isomonodromic deformations, the correct way to deduce these Hamiltonians is to express equation ({\ref{laxMA}) as a time-dependent Hamiltonian system.

\subsection{Spectral invariants}

The spectral curve $\Gamma$ of the matrix $\mathcal{A}^{(n)}$ defined by
$$
\det(\mu I-\mathcal{A}^{(n)})=0
$$
plays an important role in defining the Hamiltonians of the isomonodromic flows and of the Painlev\'e hierarchy. We now express $\Gamma$ in terms of the generating functions of the canonical coordinates and the time:
\begin{eqnarray*}
\mu^2&=&
\lambda^{-2}\left(\mathcal{A}^2-\mathcal{B}_{odd}^2+(\mathcal{B}_{even}-\alpha_n)^2\right)\\
&=&\left((2\lambda)^{4n}(1+\mathcal{P})^2-
4(2\lambda)^{2n-1}(1+\mathcal{P})\mathcal{B}_{odd}+
\lambda^{-2}(\mathcal{B}_{even}-\alpha_n)^2\right)
\end{eqnarray*}
By substituting (\ref{eq:ABB}) into the above, we obtain the
spectral curve in terms of the canonical coordinates
\begin{eqnarray*}
\mu^2&=&(2\lambda)^{4n}(1+\mathcal{P})^2-
4(2\lambda)^{2n-1}(1+\mathcal{P})\mathcal{B}_{odd}+
\lambda^{-2}(\mathcal{B}_{even}-\alpha_n)^2=\\
&=&(2\lambda)^{4n}\left(1+\mathcal{P}\right)^2-(2\lambda)^{2n-1}(1+\mathcal{P})
\Bigg[(2\lambda)^{2n+1}\left(1+\mathcal{P}-{{(1+\mathcal{T})^2}\over{1+\mathcal{P}}}\right)+\\
&+&4(2\lambda)^{-2n-1}\mathcal{Q}^2(1+\mathcal{P})\Bigg]_+
+\lambda^{-2}\left(\lambda^2\left[\lambda^{-2}\mathcal{Q}(1+\mathcal{P})\right]_+ +\alpha_n\right)^2=\\
&=&(2\lambda)^{2n-1}(1+\mathcal{P})
\Bigg[(2\lambda)^{2n+1}{{(1+\mathcal{T})^2}\over{1+\mathcal{P}}}
-4(2\lambda)^{-2n-1}\mathcal{Q}^2(1+\mathcal{P})\Bigg]_+ +\\
&+&\lambda^{-2}\left(\lambda^2\left[\lambda^{-2}\mathcal{Q}(1+\mathcal{P})\right]_+ +\alpha_n\right)^2
\end{eqnarray*}
where ${1\over{1+\mathcal{P}}}=\sum_{i=0}^\infty(-1)^i\mathcal{P}^i$. The last equality follows be $\lambda^{2n}(1+\mathcal{P})$
contains positive powers only.

In particular, we have proved the following
\begin{proposition}
The coefficients of the spectral curve $\mu^2$ of the matrix $\mathcal{A}^{(n)}$ can be expressed as polynomials of the canonical coordinates and the times as follows
\begin{eqnarray}\label{eq:mu2}
\mu^2&=&(2\lambda)^{2n-1}(1+\mathcal{P})
\Bigg[(2\lambda)^{2n+1}{{(1+\mathcal{T})^2}\over{1+\mathcal{P}}}
-4(2\lambda)^{-2n-1}\mathcal{Q}^2(1+\mathcal{P})\Bigg]_+\nonumber\\
&+&\lambda^{-2}\left(\lambda^2\left[\lambda^{-2}\mathcal{Q}(1+\mathcal{P})\right]_+-\alpha_n\right)^2.
\end{eqnarray}
\end{proposition}

\begin{corollary}\label{co:si}
The constant $\alpha_n$, the times $t_1,\dots t_{n-1}$ and the Hamiltonian $H^{(n)}$ are spectral invariants. In particular the spectral curve can be written as
$$
\mu^2 = (2\lambda)^{2n}\left(
(2\lambda)^{2n}(1+\mathcal{T})^2\right)_+ -4^n H^{(n)} \lambda^{2n-2}
+\sum_{k=1}^{n-1} h^{(n)}_k \lambda^{2k-2}
+\frac{\alpha_n^2}{\lambda^2},
$$
where
$$
h^{(n)}_k={1\over 2}{\res}\tr\left(\lambda^{1-2k}\left(\mathcal{A}^{(n)}\right)^2\right),
\qquad k=1,\dots,n.
$$
\end{corollary}

\begin{proof}
By definition of $h^{(n)}_k$, it is clear that $h^{(n)}_k$ is the coefficient of the $2k-2$ power in $\lambda$. Analogously for $H^{(n)}$. So we only need to prove that the coefficients of the powers
$4n, 4n-2,\dots,2n$ are given by $(2\lambda)^{2n}\left(
(2\lambda)^{2n}(1+\mathcal{T})^2\right)_+$. In particular, this implies that the polynomial part of $\mu$ is given by the following
\begin{eqnarray*}
\mu=4^{n}t_n\lambda^{2n}+4^{n-1}t_{n-1}\lambda^{2n-2}+\ldots-z.
\end{eqnarray*}
In fact, let the expansion of the spectrum $\mu$ at $\lambda=\infty$ be the following
\begin{eqnarray*}
\mu=\sum_{i=-\infty}^{2n}\mu_i\lambda^{i}
\end{eqnarray*}
Let us denote the coefficients of $\mu^2$ by $D_i$, that is
\begin{eqnarray*}
\mu^2=\sum_{i=-2}^{2n}D_i\lambda^i
\end{eqnarray*}
We can express $D_i$ as quadratic polynomials in the $\mu_k$ and if $k<0$, then $\mu_k$ will only appear in the coefficient $D_i$ when $i<2n$. Therefore, to compute the polynomial part of $\mu$, we only need to consider the coefficients $D_i$ with $i>2n-1$. These coefficients are given by the coefficients of
\begin{eqnarray*}
((2\lambda)^{-2n}\mu^2)_+=\left((2\lambda)^{2n}\left((2\lambda)^{-4n}\mu^2\right)\right)_+
\end{eqnarray*}
By using the relations (\ref{eq:xy2}) in (\ref{eq:mu2}), we see that
\begin{eqnarray*}
((2\lambda)^{-2n}\mu^2)_+&=&
\Bigg[(2\lambda)^{2n}{(1+\mathcal{T})^2}
-4(2\lambda)^{-2n-2}\mathcal{Q}^2(1+\mathcal{P})^2\Bigg]_+\nonumber\\
&+&\left[4(2\lambda)^{-2n-2}\mathcal{Q}^2(1+\mathcal{P})^2\right]_+\\
((2\lambda)^{-2n}\mu^2)_+&=&
\left[(2\lambda)^{2n}{(1+\mathcal{T})^2}\right]_+
\end{eqnarray*}
This implies the corollary. \end{proof}

\subsection{Time flows Hamiltonians}

We want to adapt the construction of Section \ref{sec:coaj} to express equation ({\ref{laxMA}) as a
time-dependent Hamiltonian system and the computations of Section \ref{se:ham}  to find the Hamiltonians.

The main difficulty we encounter is that now
$$
\p_{\lambda}\hat M^{(k)}\neq(2k+1)\p_{{t}_k}^w\mathcal{A}^{(n)}.
$$
The main idea to handle this problem
is the following: suppose there exists a set of coordinates $u_1,u_2,\ldots, u_{2n}$ in our
coadjoint orbit $\mathcal O_A$ such that
\begin{equation}\label{eq:u}
\p_{\lambda}\hat M^{(k)}=(2k+1)\p_{{t}_k}^u\mathcal{A}^{(n)},
\end{equation}
where $\p_{{t}_k}^u$ denotes the ${t}_k$--derivative with the $u$
coordinates fixed (in the sense explained at the beginning of Section \ref{sec:coaj}).
Then (\ref{laxMA}) becomes
\begin{equation}\label{eq:MuA}
(2k+1)\left(\p_{{t}_k}-\p_{{t}_k}^u\right)\mathcal{A}^{(n)}=[\hat M^{(k)},\mathcal{A}^{(n)}]
\end{equation}
after cancelation. This allows us to interpret the evolution along
$\p_{{t}_k}-\p_{{t}_k}^u$ as a Hamiltonian vector field on the
coadjoint orbit, as explained in Section \ref{sec:coaj}, provided
that the coordinates $u_1,\dots,u_{2n}$ exist (which is a
non--trivial fact because equation (\ref{eq:u}) is overdetermined).

Our strategy is as follows: since in Proposition \ref{prop:ham} we computed the Hamiltonians $h^{(n)}_k$ corresponding to the matrices $\hat L_k$ which give the same flow as
\begin{eqnarray}\label{eq:Lk}
L_k=\left[\lambda^{1-2k}\mathcal{A}^{(n)}\right]_+, \qquad
k=1,\ldots, n,
\end{eqnarray}
we introduce some new times $s_1,\dots,s_{n}$ corresponding to
the flows along $L_1,\dots.L_{n}$:
\begin{eqnarray}\label{eq:dsk}
\p_{s_k}\mathcal{A}^{(n)}=[L_k,\mathcal{A}^{(n)}]+\p_{\lambda}L_k
, \quad k=1,\ldots, n.
\end{eqnarray}
Observe that $s_n=-\frac{z}{4^n}$.
Since the matrices $L_k$ are related to the matrices $\hat{M}^{(k)}$ by
\begin{eqnarray}\label{eq:LM}
L_k=4^k\sum_{i=k+1}^{n}{t}_{i}\hat{M}^{(i-k)}-4^k{t}_k\mathcal{B},\qquad
k=1,\ldots, n,
\end{eqnarray}
the times ${t}_1,\dots,{t}_{n-1}$ and the $s_1,\dots,s_{n-1}$ must be related by:
\begin{eqnarray}\label{eq:st}
\p_{s_k}=4^k\sum_{i=k+1}^n(2(i-k)+1){t}_i\p_{{t}_{i-k}}-4^k{t}_k\p_z,\qquad
k=1,\ldots, n.
\end{eqnarray}
This relation is the key to our procedure: on one side, it will allow us to prove that
 there exist coordinates $u_1,\dots,u_{2n}$ such that
\begin{eqnarray}\label{eq:exder}
\p_{s_k}^u\mathcal{A}^{(n)}=\p_{\lambda}L_k,
\end{eqnarray}
on the other side relation (\ref{eq:LM}), will allow us to compute
the Hamiltonians $H^{(n)}_k$ in terms of the Hamiltonians
$h^{(n)}_k$.

The following proposition shows that (\ref{eq:st})guarantees the compatibility of the over--determined system (\ref{eq:u}).

\begin{proposition}\label{pro:Bk}
Let $s_k$ be the times corresponding to the equations
\begin{eqnarray*}
\p_{s_k}\mathcal{A}^{(n)}-\p_{\lambda}L_k=[L_k,\mathcal{A}^{(n)}]
, \quad k=1,\ldots, n
\end{eqnarray*}
where $L_k$ are given by (\ref{eq:Lk}). Then the system of differential equations
\begin{eqnarray*}
\p_{s_k}^u\mathcal{A}^{(n)}=\p_{\lambda}L_k
\end{eqnarray*}
is compatible only if ${t}_k$ and $s_k$ satisfy the following relations
\begin{eqnarray}\label{eq:sdiff1}
\p_{s_k}{t}_j=4^k(2j+1){t}_{j+k},\quad
\p_{s_k}z=-4^k{t}_k,
\end{eqnarray}
where ${t}_l=0$ for $l>n$.
\end{proposition}

Observe that  equation (\ref{eq:sdiff1}) is
equivalent to (\ref{eq:st}).

\begin{proof} Let us consider the explicit $s_k$ derivative given
by
\begin{eqnarray*}
\p_{s_k}^u\mathcal{A}^{(n)}=\p_{\lambda}L_k.
\end{eqnarray*}
This is an over--determined system of equations. We are going to prove that it admits a solution.
In terms of the entries of $\mathcal{A}^{(n)}$, this is
equivalent to
\begin{eqnarray*}
\p_{s_k}^ua_{2j+1}^{(n)}&=&(2j+1)a_{2j+2k+1}^{(n)},\quad j=0,\ldots, n,\\
\p_{s_k}^ub_{j}^{(n)}&=&jb_{j+2k},\quad j=1,\ldots, 2n,\\
a_{j}^{(n)}&=&0, \quad j>2n+1, \quad b_j^{(n)}=0,\quad j>2n.
\end{eqnarray*}
This means that
\begin{eqnarray}\label{eq:sidiff}
\lambda^{2n}\p_{s_k}^{u}(1+\mathcal
{P})&=&\left[\p_{\lambda}\left(\lambda^{2n-2k+1}(1+\mathcal{P})\right)\right]_+,\nonumber\\
\lambda^{-1}\p_{s_k}^u\mathcal{B}_{even}&=&\left[\p_{\lambda}\left(\lambda^{-2k}\mathcal{B}_{even}\right)\right]_+,\\
\lambda^{-1}\p_{s_k}^u\mathcal{B}_{odd}&=&\left[\p_{\lambda}\left(\lambda^{-2k}\mathcal{B}_{odd}\right)\right]_+,\nonumber
\end{eqnarray}
where $\mathcal{P}$, $\mathcal{B}_{even}$ and $\mathcal{B}_{odd}$
are given by (\ref{eq:pqt}) and (\ref{eq:ABB}).
We will now rewrite the expression of $\mathcal{B}_{odd}$ in
(\ref{eq:ABB}) as
\begin{eqnarray}\label{eq:bb1}
\mathcal{B}_{odd}=&{1\over 4}\Bigg[(2\lambda)^{2n+1}(1+\mathcal{P})-{{((2\lambda)^{2n+1}(1+\mathcal{T}))^2}\over
{(2\lambda)^{2n+1}(1+\mathcal{P})}}+\nonumber\\
&+{{4\mathcal{B}_{even}^2}\over{(2\lambda)^{2n+1}(1+\mathcal{P})}}
\Bigg)\Bigg]_+.
\end{eqnarray}
Then, by computing
\begin{eqnarray}
\lambda^{-1}\p_{s_k}^u\mathcal{B}_{odd}&=&\left[\p_{\lambda}\left(\lambda^{-2k}\mathcal{B}_{odd}\right)\right]_+,
\end{eqnarray}
and applying (\ref{eq:sidiff}) and (\ref{eq:bb1}), we see that, in order that  the over--determined system (\ref{eq:exder}) is compatible, the
derivative of $\mathcal{T}$ must satisfy
\begin{eqnarray*}
\lambda^{2n}\p_{s_k}^{u}(1+\mathcal
{T})&=&\left[\p_{\lambda}\left(\lambda^{2n-2k+1}(1+\mathcal{T})\right)\right]_+.
\end{eqnarray*}
This gives
\begin{eqnarray}\label{eq:sdiff}
\p_{s_k}^{u}{t}_j=4^k(2j+1){t}_{j+k},\quad
\p_{s_k}^uz=-4^k{t}_k,\nonumber
\end{eqnarray}
where ${t}_{l}=0$ if $l>n$ in the above equation.

If we chose another set of coordinates,
\begin{eqnarray*}
\{y_1(u,t),\ldots, y_{2n}(u,t), {t}_1,\ldots, {t}_n\}
\end{eqnarray*}
Then the vector field $\p_{s_k}^{y}$ is given by
\begin{eqnarray*}
\p_{s_k}^{u}=\p_{s_k}^{y}+\sum_{j=1}^n\p_{s_k}^uy_j \p_{y_j}
\end{eqnarray*}
Therefore $\p_{s_k}^{u}t_j=\p_{s_k}^yt_j$ for $j=1,\ldots, n$ and
the same for $z$. Hence we can drop the superscript $u$ in
(\ref{eq:sdiff}) to obtain (\ref{eq:sdiff1}).
\end{proof}

As proved in Proposition \ref{prop:ham}, the Hamiltonians $h^{(n)}_k$ corresponding to the vector field
$$
\mathcal X_k:=[L_k,A],
$$
are given by equation (\ref{eq:ham1}). Due to our derivation, they
must be thought of as written in the coordinates $u$. Since the canonical coordinates $P_1,\ldots,Q_n$ depends explicitly on the $s_k$ when expressed in terms of the coordinates $u$, the actual Hamiltonian differs by a shift given by Lemma \ref{le:bo}.

We first compute the explicit $s_k$ derivatives of the canonical coordinates. From the definition of the coordinates $Q_1,\dots,Q_n$ we have
\begin{eqnarray*}
\lambda^{-2}\mathcal{Q}=-\left[\lambda^{-2}\mathcal{B}_{even}(1+\mathcal{P})^{-1}\right]_+.
\end{eqnarray*}
We can the apply (\ref{eq:sidiff}) to compute the explicit $s_i$
derivatives of $\mathcal{Q}$. This gives
\begin{eqnarray*}
\p_{s_k}^u\mathcal{Q}=\left[\lambda^{1-2k}\p_{\lambda}\mathcal{Q}-(2n+1)\lambda^{-2k}\mathcal{Q}\right]_+.
\end{eqnarray*}
Hence the explicit $s_j$ derivatives of the canonical coordinates
are given by
\begin{eqnarray}\label{eq:dspq}
\p_{s_k}^uQ_j&=&(2(j+k)-2n-1)Q_{j+k},\nn\\
\p_{s_k}^uP_j&=&(2(n-j)+1)P_{j-k},\quad j\neq k\\
\p_{s_k}^uP_k&=&(2(n-k)+1),\nn
\end{eqnarray}
where if $l>n$ or $l<1$, then $P_l=0$, $Q_l=0$.

Using Lemma \ref{le:bo} and equation  (\ref{eq:dspq}),  we can compute the shifts :
\begin{eqnarray}\label{eq:dhs}
\delta h^{(n)}_{k}=\sum_{j=1}^{n-k}(2(j+k)-2n-1)P_jQ_{j+k}+(2k-2n-1)Q_k.
\end{eqnarray}

Finally, we use the formulae above to show that the coordinates $u$ actually exist. In fact  we construct them recursively as follows using (\ref{eq:dspq}). For $j=1$ this gives
$$
\p_{s_k}^uP_1=(2n-1)\delta_{k1},
$$
so that we choose $u_1=P_1-(2n-1)s_1$. Then for $j=2$ equation  (\ref{eq:dspq}) gives:
\begin{eqnarray*}
\p_{s_1}^u P_2&=&(2(n-2)+1)P_1,\\
\p_{s_2}^uP_2&=&(2n-1)
\end{eqnarray*}
and all other derivatives of $P_2$ are zero, so we put
$$
u_2=
P_2-(2n-1)s_2-(2(n-2)+1)P_1s_1+(2(n-2)+1)(2n-1){s_1^2\over 2},
$$
so that $p_{s_k}^u u_2=0$. By repeating this procedure iteratively we get:
\begin{eqnarray*}
u_{i}&=&P_i+\sum_{j=1}^{i-1}W_{ij}P_{i-j}+W_{ii},\quad i=1,\ldots, n,\\
u_{i+n}&=&Q_i+\sum_{j=1}^{n-i}V_{ij}Q_{i+j},\quad i=1,\ldots, n,
\end{eqnarray*}
where $W_{ij}$ and $V_{ij}$ are given by
\begin{eqnarray*}
W_{ij}&=&\sum_{m=1}^{j}\sum(-2(n-i+j-l_1)-1)^{k_1}\left({s_{l_1}}\over{k_1}\right)^{k_1}
\cdots(-2(n-i+j-l_c)-1)^{k_c}\left({s_{l_c}}\over{k_c}\right)^{k_c},\\
V_{ij}&=&\sum_{m=1}^{j}\sum(-2(i+j-l_1)+2n+1)^{k_1}\left({s_{l_1}}\over{k_1}\right)^{k_1}
\cdots(-2(i+j-l_c)+2n+1)^{k_c}\left({s_{l_c}}\over{k_c}\right)^{k_c},
\end{eqnarray*}
where the second summations in the above are taken over all possible combinations of integers
\begin{eqnarray*}
l_1k_1&+&l_2k_2+\cdots+l_ck_c=j \\
k_1&+&\cdots+k_c=m,\quad 0<k_i<m, \quad 0<l_i<j.
\end{eqnarray*}

Resuming, we proved the following:

\begin{theorem}\label{thm:hamsk}
The Hamiltonians for the equations
\begin{eqnarray}\label{eq:lk}
\p_{s_k}\mathcal{A}^{(n)}-\p_{\lambda}L_k=[L_k,\mathcal{A}^{(n)}]
\end{eqnarray}
where $L_k$ are defined by (\ref{eq:Lk}), are
\begin{eqnarray}\label{eq:hs}
h^{(n)}_k={1\over
2}{\res}\tr\left(\lambda^{1-2k}\left(\mathcal{A}^{(n)}\right)^2\right)
\end{eqnarray}
and their shifts $\delta h^{(n)}_k$ corresponding to the canonical
coordinates $P_k$, $Q_k$ defined in Theorem \ref{mainth1} are given by
\begin{eqnarray*}
\delta h^{(n)}_k=\sum_{j=1}^{n-k}(2(j+k)-2n-1)P_jQ_{j+k}+(2k-2n-1)Q_k,
\end{eqnarray*}
that is, the equations (\ref{eq:lk}) can be expressed as the
following time-dependent Hamiltonian equations
\begin{eqnarray*}
\p_{s_k}\mathcal{A}^{(n)}=\left\{h^{(n)}_k+\delta
h^{(n)}_k,\mathcal{A}^{(n)}\right\}+\p_{s_k}^{can}\mathcal{A}^{(n)}
\end{eqnarray*}
with the Poisson bracket $\{,\}$ defined by
\begin{eqnarray}\label{eq:bracket1}
\{f,g\}=\sum_{i=1}^n{{\p f}\over{\p P_i}}{{\p g}\over{\p
Q_i}}-{{\p g}\over{\p P_i}}{{\p f}\over{\p Q_i}},
\end{eqnarray}
and the derivative $\p_{s_k}^{can}$ is the explicit time
derivative of $s_k$ when the canonical coordinates $P_i$, $Q_i$
are fixed.
\end{theorem}

Now let us deal with the ${t}_k$--flows. Thanks to the fact that the coordinates $u_1,\dots, u_{2n}$ exist and are such that (\ref{eq:exder}) is satisfied, equations  (\ref{eq:LM}) and (\ref{eq:st}) we see that the
matrices $\hat{M}^{(i)}$ correspond to the times
$(2i+1){t}_i$. That is, the equations
\begin{eqnarray}\nn
(2k+1)\p_{{t}_k}^u\mathcal{A}^{(n)}&=&\p_{\lambda}\hat{M}^{(k)}\\
(2k+1)\p_{{t}_k}\mathcal{A}^{(n)}&=&[\hat{M}^{(k)},\mathcal{A}^{(n)}]+\p_{\lambda}\hat{M}^{(k)}
, \quad k=1,\ldots, n\nn
\end{eqnarray}
are compatible and hence we can express the higher order isomonodromic flows (\ref{timeflows}) as time-dependent Hamiltonian flows.

We can now compute the Hamiltonians $H^{(n)}_k$ for the times
${t}_k$ by using Theorem \ref{thm:hamsk} and (\ref{eq:st}).

\begin{corollary}\label{co:hamt}
Let  $\mathcal K_s$  and  $\mathcal H_{\hat t}$ be the following polynomials in $\lambda$ and $\frac{1}{\lambda}$ respectively:
$$
\mathcal{K}_{s}=\sum_{k=1}^{n}(h^{(n)}_k+ \delta
h^{(n)}_k)\lambda^{2k}, \qquad
\mathcal{H}_{{t}}=\sum_{k=1}^{n-1}(H^{(n)}_k+ \delta
H^{(n)}_k)(2\lambda)^{-2k-1}- \frac{1}{2\lambda}(H^{(n)}+\delta
H^{(n)}),
$$
where $h^{(n)}_k$, $\delta h^{(n)}_k$ are given by (\ref{eq:hs}) and (\ref{eq:dhs}) respectively and
the coefficients $H^{(n)}_k+\delta H^{(n)}_k$ are given  by:
\begin{eqnarray}\label{relHK}
\p_{\lambda}\mathcal{H}_{{t}}&=& -2(2\lambda)^{-2n}\left(
(2\lambda)^{-2}\mathcal{K}_s(1+\mathcal{T})^{-1}\right)_+,
\end{eqnarray}
with $\mathcal{T}$ is defined by (\ref{eq:pqt}).
Then the equations
\begin{eqnarray*}
(2k+1)\p_{{t}_k}\mathcal{A}^{(n)}-\p_{\lambda}\hat{M}^{(k)}=[\hat{M}^{(k)},\mathcal{A}^{(n)}]
\end{eqnarray*}
can be expressed as time-dependent Hamiltonian equations with the
Poisson bracket given by (\ref{eq:bracket1}) as follows:
\begin{eqnarray*}
\p_{{t}_k}\mathcal{A}^{(n)}=\left\{H^{(n)}_k+\delta
H^{(n)}_k,\mathcal{A}^{(n)}\right\}+\p_{{t}_k}^{can}\mathcal{A}^{(n)},
\end{eqnarray*}
where the derivative $\p_{{t}_k}^{can}$ is the explicit time ${t}_k$
derivative when the canonical coordinates $P_i$,
$Q_i$ are fixed. In particular, the equation (\ref{timeflows})
is Hamiltonian and it is equivalent to
$$
\frac{\partial Q_i}{\partial {t}_k} = \frac{\partial\mathcal H^{(n)}_k}{\partial P_i},\quad
\frac{\partial P_i}{\partial {t}_k} = -\frac{\partial\mathcal H^{(n)}_k}{\partial Q_i},
\qquad k=1,\dots,n-1,\quad  i=1,\dots,n,
$$
where
$$
\mathcal H^{(n)}_k= H^{(n)}_k+\delta H^{(n)}_k,\qquad k=1,\dots, n-1.
$$
\end{corollary}

\begin{proof} From (\ref{eq:st}), we see that the Hamiltonians are
related by
\begin{eqnarray*}
h^{(n)}_k+\delta
h^{(n)}_k=4^k\sum_{i=k+1}^n(2(i-k)+1){t}_i(H^{(n)}_{i-k}+\delta
H^{(n)}_{i-k}) -4^k{t}_k(H^{(n)}+\delta H^{(n)}).
\end{eqnarray*}
This implies the following relation
\begin{eqnarray*}
2\mathcal{K}_s&=&-(2\lambda)^2\left((2\lambda)^{2n}\p_{\lambda}\mathcal{H}_t(1+\mathcal{T})\right),\\
\left((2\lambda)^{2n}(2(2\lambda)^{-2n-2}\mathcal{K}_s)\right)_+&=&-\left((2\lambda)^{2n}\p_{\lambda}\mathcal{H}_t(1+\mathcal{T})\right).
\end{eqnarray*}
Then, by using (\ref{eq:xy2}), we obtain (\ref{relHK}).
\end{proof}

\begin{example}
Let us demonstrate the results of this Section in the first non-trivial example: $n=2$.
We want to represent the ${t}_1$ flow
\begin{eqnarray}\label{eq:t1}
3\p_{{t}_1}\mathcal{A}^{(2)}-\p_{\lambda}\hat{M}^{(1)}=[\hat{M}^{(1)},\mathcal{A}^{(2)}]
\end{eqnarray}
where
\begin{eqnarray*}
\hat{M}^{(1)}=\begin{pmatrix}
4\lambda^{3}-2w^2\lambda&-4w\lambda^2+2w_z\lambda-w_{zz}+2w^3\\
                       -4w\lambda^2-2w_z\lambda-w_{zz}+2 w^3&-4\lambda^{3}+2w^2\lambda\\
                       \end{pmatrix},
\end{eqnarray*}
\begin{eqnarray}\nn
&&
\mathcal A^{(2)} =\left(\begin{array}{c}
16\lambda^4 + 4 \lambda^2(t_1-2 w^2) -z -2 t_1 w^2+6 w^4 +2 w_z^2-4 w w_{zz}\\
-16 \lambda^3 w-8 \lambda^2 w_z + 4 \lambda (2 w^3-w t_1- w_{zz})
- 2t_1 w_z+12 w^2 w_z -\frac{\alpha_2}{\lambda}\\
\end{array}\qquad\qquad\qquad\qquad\qquad\qquad\qquad\qquad\right.\nn\\
&&\qquad\qquad\qquad\qquad\qquad\qquad
\left.\begin{array}{c}
-16 \lambda^3 w+8 \lambda^2 w_z + 4 \lambda (2 w^3-w t_1- w_{zz})
+ 2t_1 w_z-12 w^2 w_z -\frac{\alpha_2}{\lambda}\\
-16\lambda^4 - 4 \lambda^2(t_1-2 w^2) +z +2 t_1 w^2-6 w^4 -2 w_z^2+4 w w_{zz}
                       \end{array}\right)\nn
\end{eqnarray}
as a time dependent Hamiltonian equation.

We are now going to show that there exist coordinates $u_1,\dots,u_4$ such that
\begin{eqnarray*}
\p_{\lambda}\hat{M}^{(1)}=3\p_{{t}_1}^u\mathcal{A}^{(2)}.
\end{eqnarray*}
This gives the following equations
\begin{subequations}\nn
\begin{eqnarray}
3\p_{{t}_1}^ua_3^{(2)}&=&3\p_{{t}_1}^u\left(-8w^2+4{t}_1\right)=
12,\label{eq:a3}\\
3\p_{{t}_1}^ua_1^{(2)}&=&
3\p_{{t}_1}^u\left(2(w_z^2+3w^4-2ww_{zz})-2{t}_1w^2-z\right)=-2w^2,\label{eq:a1}\\
3\p_{{t}_1}^ub_4^{(2)}&=&-48\p_{{t}_1}^uw=0,\label{eq:b4}\\
3\p_{{t}_1}^ub_3^{(2)}&=&24\p_{{t}_1}^uw_z=0,\label{eq:b3}\\
3\p_{{t}_1}^ub_2^{(2)}&=&3\p_{{t}_1}^u\left(-4w_{zz}+8w^3-4{t}_1w\right)=
-8w,\label{eq:b2}\\
3\p_{{t}_1}^ub_1^{(2)}&=&3\p_{{t}_1}^u\left(2(w_{zzz}-6w^2w_z)+2{t}_1w_z\right)=2w_z.\label{eq:b1}
\end{eqnarray}
\end{subequations}
We also need $\p_{{t}_1}^u{t}_1=1$ and
$\p_{{t}_1}^uz=0$. Of course, after imposing these two
constraints, we have more equations then independent variables. (6
equations and 4 independent variables $w$, $w_z$, $w_{zz}$ and
$w_{zzz}$) We need to check that equations
(\ref{eq:a3})-(\ref{eq:b1}) are consistent. To see this, first note
that equations (\ref{eq:b4}) and (\ref{eq:b3}) imply
\begin{eqnarray*}
\p_{{t}_1}^uw=\p_{{t}_1}^uw_z=0.
\end{eqnarray*}
By substituting this into (\ref{eq:a3}), we see that
\begin{eqnarray*}
3\p_{{t}_1}^ua_3^{(2)}=3\p_{{t}_1}^u\left(-8w^2+4{t}_1\right)=12,
\end{eqnarray*}
which is tautologically true, so (\ref{eq:a3}) is consistent with
(\ref{eq:b3}) and (\ref{eq:b4}). By substituting (\ref{eq:b3}) and
(\ref{eq:b4}) into (\ref{eq:a1}), we see that
\begin{eqnarray*}
3\p_{{t}_1}^ua_1^{(2)}&=&3\p_{{t}_1}^u\left(2(w_z^2+3w^4-2ww_{zz})-2{t}_1w^2-z\right)=-2w^2\\
&\Rightarrow&\p_{{t}_1}^uw_{zz}=-{1\over 3}w.
\end{eqnarray*}
Then by substituting these into (\ref{eq:b2}), we find that
\begin{eqnarray*}
3\p_{{t}_1}^ub_2^{(2)}&=&3\p_{{t}_1}^u\left(-4w_{zz}+8w^3-4{t}_1w\right)\\
&=&-12\p_{{t}_1}^uw_{zz}-12w=-8w,
\end{eqnarray*}
therefore (\ref{eq:b2}) is also consistent with the other
equations. Now the last equation (\ref{eq:b1}) does not have any
consistency issue and it would give us the following
\begin{eqnarray*}
3\p_{{t}_1}^uw_{zzz}=-2w_z.
\end{eqnarray*}
Therefore these equations are all consistent and $u_1,\dots,u_4$ exist.

Using the results of Section \ref{se:expl}, we can express the matrix $\mathcal A^{(n)}$ in terms of the canonical coordinates $P_1,P_2,Q_1,Q_2$ and using formula (\ref{eq:hs}) we get
$$
h^{(2)}_1 =- 32 z P_2 +256 P_1^2 P_2-256 P_2^2-2 P_2 Q_1 Q_2- P_1 P_2 Q_2^2
-2  \alpha_2(P_1 Q_2-Q_1)- t_1P_2 (128 P_1-16 t_1)
$$
$$
h^{(2)}_2=-32 z P_1 +256 P_1^3 -512 P_1 P_2 -P_2 Q_2^2+
Q_1^2+2 \alpha_2 Q_2 -
 {t}_1 \left( 128 P_1^2-128 P_2-16 P_1 {t}_1 \right),
$$
and using (\ref{eq:dhs}) we get for the shifts:
$$
\delta h^{(2)}_1=-P_1 Q_2-Q_1,\qquad
\delta h^{(2)}_2= -Q_2,
$$
so that, using (\ref{relHK}), the Hamiltonian of the equation ${\rm P_{II}}^{(2)}$ is
$$
{\mathcal H}^{(2)} = \frac{1}{16} (-h^{(2)}_1+Q_2).
$$
and the Hamiltonian of (\ref{timeflows}) with $k=1$ is
$$
\mathcal H^{(2)}_1= \frac{1}{3} \left(\frac{1}{4}(h^{(2)}_1+\delta h^{(2)}_1)-\frac{1}{16} t_1 (h^{(2)}_2+\delta h^{(2)}_2)\right).
$$
For the sake of completeness, we can compute the shift in $\mathcal H^{(2)}_1$  directly from the explicit dependence of $P_1,P_2,Q_1,Q_2$ on $s_1$. From (\ref{eq:a3})-(\ref{eq:b1}), we have
\begin{eqnarray*}
\p_{{t}_1}^uP_1&=&{1\over
16}\p_{{t}_1}^u(a_3^{(2)}+b_3^{(2)})={1\over 4},\\
\p_{{t}_1}^uP_2&=&{1\over
16}\p_{{t}_1}^u(a_1^{(2)}+b_1^{(2)})={1\over 24}{\mathcal L}_1={P_1\over 12}-{{t}_1\over 48},\\
\p_{{t}_1}^uQ_2&=&0,\\
\p_{{t}_1}^uQ_1&=&-\p_{{t}_1}^ub_2^{(2)}-Q_2\p_{{t}_1}^uP_1
=-{Q_2\over 12}.
\end{eqnarray*}
Therefore the we can find $\delta H^{(2)}_1$ from the following partial
differential equations
\begin{eqnarray*}
{{\p\delta H^{(2)}_1}\over{\p Q_1}}&=&-{1\over 4},\\
{{\p\delta H^{(2)}_1}\over{\p Q_2}}&=&-{P_1\over 12}+{{t}_1\over 48},\\
{{\p\delta H^{(2)}_1}\over{\p P_1}}&=&-{Q_2\over 12},\\
{{\p\delta H^{(2)}_1}\over{\p P_2}}&=&0.
\end{eqnarray*}
This gives the shift
\begin{eqnarray*}
\delta H^{(2)}_1=-{Q_1\over 4}-{{P_1Q_2}\over 12}+{{{t}_1Q_2}\over
{48}},
\end{eqnarray*}
which in fact agrees with what we obtained above.
\end{example}

\appendix

\section{From the mKdV Lax pair to the isomonodromic problem}\label{app-a}

Let us consider the Lax pair of the mKdV hierarchy:
$$
\frac{\partial\Phi}{\partial x } = N\Phi,
\qquad
\frac{\partial\Phi}{\partial t_{k+1} } = N_k\Phi,
$$
where
$$
N=\left(\begin{array}{cc}
-\zeta& v\\
v& \zeta
\end{array}\right),\qquad
N_k=\left(\begin{array}{cc}
       \sum_{j=1}^{2k+1}{A}_{j}^{(k)}\zeta^{j}&
\sum_{j=0}^{2l}{B}_{j}^{(k)}\zeta^{j}\\
       \sum_{j=0}^{2l}{C}_{j}^{(k)}\zeta^{j}
&-\sum_{j=1}^{2l+1}{A}_{j}^{(k)}\zeta^{j}
\end{array}\right),
$$
where the coefficients  $A_j^{(k)},B_j^{(k)},C_j^{(k)}$ are given by
the formulae (\ref{isomonodromy.4}) with ${\mathcal L}_l$
replaced by  $\mathcal R_l$,  $z$ replaced by $x$ and $w$ replaced
by $v$. Observe that in $N_k$ the terms $B_0^{(k)}$ are nonzero.
They are
$$
B_0^{(k)}=-(\partial_x+2 v){\mathcal R}_k.
$$
Details can be found in \cite{refCJM}.

By the self--similarity reduction (\ref{self-sim}), and defining
$$
\lambda=[(2n+1)T_{n+1} ]^{\frac{1}{2n+1}} \zeta,
$$
we obtain:
$$
(2n+1)T_{n+1}\frac{\partial\Psi}{\partial T_{n+1} } =-z\frac{\partial \Psi}{\partial z} -
\sum_{1}^{n-1}(2k+1){t}_k \frac{\partial \Psi}{\partial {t}_k} +
\lambda\frac{\partial\Psi}{\partial\lambda},
$$
$$
T_{k+1}\frac{\partial\Psi}{\partial T_{k+1} } ={t}_{k}\frac{\partial\Psi}{\partial{t}_{k} } ,
$$
$$
x\frac{\partial\Psi}{\partial x } =z\frac{\partial\Psi}{\partial z}.
$$
Now, defining
$$
\mathcal B = [(2n+1)T_{n+1} ]^{\frac{1}{2n+1}} N,\qquad
\hat M^{(k)}=[(2n +1)T_{n+1}]^{\frac{2k+1}{2n+1}} N_k,\quad k=1,\dots,n,
$$
we get
\begin{subequations}\label{is-red}
\begin{eqnarray}
{\partial\Psi\over\partial z}&=&{\mathcal{B}}\Psi,\\
{\partial\Psi\over\partial\lambda}&=&
\frac{1}{\lambda}\left[
\hat M^{(n)}+\sum_{l=1}^{n-1} {t}_{l} \hat M^{(l)} +z \,{\mathcal B}
\right]\Psi,\\
(2k+1)\frac{\p\Psi}{\p {t}_k}&=& \hat M^{(k)} \Psi.
\end{eqnarray}
\end{subequations}
To show that (\ref{is-red}) coincide with (\ref{isomonodromy}) first observe that
\begin{equation}\label{betta1}
\hat M_k=M^{(k)} - \left(
\begin{array}{cc} 0&(\partial_z+2 w){\mathcal L}_k\\
(\partial_z+2 w){\mathcal L}_k&0
\end{array}\right),
\end{equation}
as it follows from the definition of $\hat M_k$ and $N_k$. Moreover
\begin{equation}\label{betta}
{\mathcal A}^{(n)}=\hat M^{(n)}+\sum_{l=1}^{n-1} {t}_{l} \hat M^{(l)} +z \,{\mathcal B},
\end{equation}
because
$$
-(\partial_z+2 w){\mathcal L}_n-\sum_{l}^{n-1}
{t}_{l}  (\partial_z+2 w){\mathcal L}_l+z w=-\alpha_n
$$
thanks to (\ref{timepii}).

It is now clear how to prove that the equations (\ref{comp-ii}), (\ref{eq-MB}) and (\ref{laxMA}) are indeed consistent. This follows from the fact that the mKdV flows commute, so that
$$
\frac{\partial N}{\partial T_{k+1}} - \frac{\partial N_k}{\partial x } N_k=[N_k,N],
$$
and
$$
\frac{\partial N_l}{\partial T_{k+1}} - \frac{\partial N_k}{\partial T_{l+1} } =[N_k,N_l].
$$
The equations (\ref{comp-ii}), (\ref{eq-MB}) and (\ref{laxMA}) then follow automatically from the relations
(\ref{betta}), (\ref{betta1}).


\def\refjl#1#2#3#4#5#6#7{\bibitem{#1}{\frenchspacing\rm#2}, #6,
{\frenchspacing\it#3\/}, {\bf#4} (#7) #5.}

\def\refbk#1#2#3#4#5{\bibitem{#1}{\frenchspacing\rm#2},
``{\sl#3\/}'', #4 (#5).}

\def\refpp#1#2#3#4#5{
\bibitem{#1}
\smallskip\noindent{\frenchspacing\rm#2}, {``#3''}\ #4\ (#5).}

\def\refcf#1#2#3#4#5#6#7{
\bibitem{#1}
\smallskip\noindent{\frenchspacing\rm#2}, {\rm#3},
in ``{\sl#4\/}'' [#5] #6\ (#7).}

\def\sideiii#1#2#3#4{\refcf{#1}{#2}{#3}{SIDE III --- Symmetries and
Integrability of Difference  Equations}{D.\ Levi and O.\ Ragnisco,
Editors}{{\frenchspacing\it CRM Proc. Lect. Notes Series\/}, vol.
{\bf25}, Amer. Math. Soc., Providence, RI, pp.\ #4}{2000}}

\end{document}